\documentclass[thmsa,12pt]{article}
\usepackage{amsfonts}

\usepackage{amsmath}

\input{tcilatex}

\begin{document}

\setcounter{page}{0} \topmargin0pt \oddsidemargin5mm \renewcommand{%
\thefootnote}{\fnsymbol{footnote}} \newpage \setcounter{page}{0} 
\begin{titlepage}
\begin{flushright}
YITP-SB-02-17 \\
\end{flushright}
\vspace{0.5cm}
\begin{center}
{\Large {\bf Universal amplitude ratios and Coxeter geometry\\
 in the dilute $A_L$ model} }

\vspace{0.8cm}
{ \large C. Korff$^1$ and K. A. Seaton$^2$}

\vspace{0.5cm}
{\em $^1$C.N. Yang Institute for Theoretical Physics\\
State University of New York at Stony Brook\\ 
Stony Brook, N.Y. 11794-3840, USA\\
\bigskip
$^2$School of Mathematical and Statistical Sciences\\
La Trobe University, Victoria 3086, Australia}
\end{center}
\vspace{0.2cm}
 
\renewcommand{\thefootnote}{\arabic{footnote}}
\setcounter{footnote}{0}

\begin{abstract}
The leading excitations of the dilute $A_L$ model in regime 2 are considered using 
analytic arguments. The model can be identified with the integrable $\phi_{1,2}$ perturbation
of  the unitary minimal series $M_{L,L+1}$. It is demonstrated that the excitation spectrum 
of the transfer matrix satisfies the same functional equations in terms of elliptic functions as 
the exact S-matrices of the $\phi_{1,2}$ perturbation do in terms of trigonometric functions. In 
particular, the bootstrap equation 
corresponding to a self-fusing process is recovered. For the special cases $L=3,4,6$ 
corresponding to the Ising model in a magnetic field, and the leading thermal perturbations of the 
tricritical Ising and three-state Potts model, as well as for the unrestricted model, $L=\infty$, we relate the 
structure of the Bethe roots to the Lie algebras $E_{8,7,6}$ and $D_4$ using Coxeter geometry. 
In these cases Coxeter geometry also allows for a single formula in generic Lie algebraic terms describing 
all four cases. For general $L$ we calculate the spectral gaps associated with the leading excitation which 
allows us to compute universal amplitude ratios characteristic of the universality class. The ratios are
 of field theoretic importance as they enter the bulk vacuum expectation value of the 
energy momentum tensor associated with the corresponding integrable quantum field theories.
\medskip
\par\noindent
\end{abstract}
\vfill{ \hspace*{-9mm}
\begin{tabular}{l}
\rule{6 cm}{0.05 mm}\\
korff@insti.physics.sunysb.edu \\
k.seaton@latrobe.edu.au
\end{tabular}}
\end{titlepage}
\newpage

\section{Introduction}

One of the most significant applications of two-dimensional integrable
systems in field theory and statistical mechanics is to systems undergoing a
continuous phase transition. Near the critical point the correlation length
diverges rendering the system scale invariant. This leads to the well-known
concept of universality classes, which describe the macroscopic scaling
behavior of the thermodynamic quantities near the critical point. The
universality classes can be identified with conformal field theories which
encode the critical exponents in the spectrum of their primary fields.
Besides the critical exponents, each class exhibits additional universal
ratios of critical amplitudes, which might equally well serve as a
characteristic, see e.g. \cite{PHA} and references therein. In two
dimensions exact non-perturbative results are available on these critical
amplitudes, provided the off-critical model can effectively be described by
an integrable perturbation of the conformal field theory governing the
critical point. There are various ways to proceed, such as conformal
perturbation theory, the bootstrap approach involving the construction of
exact scattering matrices and the transfer matrix method. While we
concentrate on the last of these in this article, it has intimate ties with
the other techniques which we will point out in the course of our
discussion.\smallskip

We consider the integrable off-critical dilute $A_{L}$ lattice model \cite
{WNS} which belongs to the class of interaction-round-a-face (IRF) models
with adjacency matrix related to the simple Lie algebra $A_{L}$. The epithet
``dilute'' points out that neighbouring lattice sites are allowed to have
the same height value in addition to those permitted in the conventional IRF
models, where neighbouring heights take adjacent values in the Dynkin
diagram. The dilute $A_{L}$ model possesses four different branches \cite
{WPSN} listed in Table 1. We focus in this article on regime 2 which can be
associated with the integrable perturbation $\phi _{1,2}$ of the unitary
minimal conformal models $M_{L,L+1}$. The Boltzmann weights in the
off-critical region are expressed in Jacobi's theta functions depending on a
spectral parameter $u$ restricted to the interval listed in Table 1, and an
elliptic nome $0<|p|<1$ which can be identified with the ordering field. The
latter is of magnetic or thermal type depending on whether $L$ is odd or
even. As $p\rightarrow 0^{\pm }$ the system approaches criticality.\medskip

\begin{center}
\begin{tabular}{|l|l|l|l|l|}
\hline\hline
regime & interval & $c$ & $\Delta ^{-1}$ & $\lambda $ \\ \hline\hline
1 & $0<u<3\lambda $ & $1-\frac{6}{(L+1)(L+2)}$ & $4\frac{L+1}{L+4}$ & $\frac{%
\pi }{4}\frac{L}{L+1}$ \\ \hline\hline
2 & $0<u<3\lambda $ & $1-\frac{6}{L(L+1)}$ & $4\frac{L+1}{L-2}$ & $\frac{\pi 
}{4}\frac{L+2}{L+1}$ \\ \hline\hline
3 & $3\lambda -\pi <u<0$ & $\frac{3}{2}-\frac{6}{(L+1)(L+2)}$ & $4\frac{L+1}{%
L-2}$ & $\frac{\pi }{4}\frac{L+2}{L+1}$ \\ \hline\hline
4 & $3\lambda -\pi <u<0$ & $\frac{3}{2}-\frac{6}{L(L+1)}$ & $4\frac{L+1}{L+4}
$ & $\frac{\pi }{4}\frac{L}{L+1}$ \\ \hline\hline
\end{tabular}
\smallskip
\end{center}

{\small \noindent \textbf{Table 1.} The four branches of the dilute A
models. Listed are the allowed values of the spectral and the crossing
parameter entering the Boltzmann weights as well as the central charge and
the anomalous scaling dimension of the perturbing operator.}\medskip\ 

In this article we put forward the leading off-critical excitations in the
spectrum of the associated transfer matrix in regime 2 for arbitrary $L$
when the thermodynamic limit is taken. This allows us to determine the
associated spectral gaps in the bulk limit, which determine the finite
off-critical correlation length 
\begin{equation}
\xi ^{-1}=-\lim_{N\rightarrow \infty }\ln \frac{\Lambda _{1}}{\Lambda _{0}}%
\sim |p|^{\frac{1}{2-2\Delta }}\,\mathcal{S}^{\pm }(q^{\prime }/|p|^{\frac{%
1-\Delta ^{\prime }}{1-\Delta }})^{-1}\;,\quad (p\rightarrow 0^{\pm })\;.
\label{xi}
\end{equation}
Here $\Lambda _{1},\Lambda _{0}$ are eigenvalues of the transfer matrix on a
strip of length $N$ corresponding to the lowest excited and the ground
state, respectively. As the critical point is approached the spectral gap
collapses, and the associated correlation length diverges with critical
exponent determined by the anomalous scaling dimension $\Delta =\Delta
_{1,2} $ of the perturbing field $\phi _{1,2}$. The renormalization group
behaviour of the system is encoded in the bulk universal scaling functions $%
\mathcal{S}^{\pm }$ which depend on the dimensionless scaling variable $%
q^{\prime }/|p|^{\frac{1-\Delta ^{\prime }}{1-\Delta }},$ with $q^{\prime }$
being a dummy variable for the perturbation by any other relevant field with
scaling dimension $\Delta ^{\prime }$. Further information on the
renormalization group characteristics of the $\phi _{1,2}$ perturbation
comes from the singular part $f_{s}$ of the free energy density of the
dilute $A_{L}$ model \cite{WPSN,BS2} and its associated bulk scaling
functions $\mathcal{Q}^{\pm } $ as $p\rightarrow 0^{\pm },$ 
\begin{equation}
f=-\lim_{N\rightarrow \infty }N^{-1}\ln \Lambda _{0}=f_{s}+f_{ns}\;,\quad
f_{s}\sim |p|^{\frac{1}{1-\Delta _{1,2}}}\,\mathcal{Q}^{\pm }(q^{\prime
}/|p|^{\frac{1-\Delta ^{\prime }}{1-\Delta }}).  \label{free}
\end{equation}
The explicit form of the bulk scaling functions $\mathcal{S}^{\pm }$ and $%
\mathcal{Q}^{\pm }$ is not accessible via the dilute $A_{L}$ model since it
realizes perturbation by only one relevant field, i.e. $q^{\prime }=0$.
However, besides the scaling dimension of the perturbing operator one can
extract the following combination of critical amplitudes (see e.g. \cite{PHA}%
): 
\begin{equation}
\lim_{p\rightarrow 0^{\pm }}\,f_{s}\cdot \xi ^{2}=\mathcal{Q}^{\pm }(0)%
\mathcal{S}^{\pm }(0)^{2}\;.  \label{amp}
\end{equation}
While each amplitude separately depends on the normalization of the fields
and is a non-universal quantity, the above combination is universal in the
sense that it only depends on the universality class determined by the
minimal model $M_{L,L+1}$ and its $\phi _{1,2}$ perturbation. Since in the
thermodynamic limit the statistical lattice model can be effectively
described by an integrable quantum field theory the amplitude (\ref{amp}) is
also accessible by exact S-matrix theory.\smallskip

The best known example of the exact S-matrices corresponding to the $\phi
_{1,2}$ perturbation of $M_{L,L+1}$ is Zamolodchikov's construction \cite{Zb}
related to the simple Lie algebra $E_{8}$ for the Ising model at $T=T_{c}$
in a magnetic field. In the context of the dilute $A_{L}$ model this case
corresponds to $L=3$ in regime 2. Further examples exhibiting the
exceptional algebraic structures $E_{7}$ and $E_{6}$ are the scattering
matrices \cite{MC,FZ,SZ,BCDS} for $L=4$ and $L=6$\ which are associated with
the leading thermal perturbation of the tricritical Ising and tricritical
three-state Potts model, respectively (see Table 2). For general $L$ it has
been proposed by Smirnov \cite{Sm} that the $\phi _{2,1}$ and $\phi _{1,2}$
perturbations should be considered as reductions of the
Zhiber-Mikhailov-Shabat (ZMS) model \cite{ZhSh,Mik}. While for arbitrary
values of the purely imaginary coupling constant the theory violates
unitarity, one can recover physically sensible theories at rational values
by reducing the state space. The corresponding scattering amplitudes are
obtained at roots of unity, $q^{L}=1,$ from the Izergin-Korepin R-matrix 
\cite{IK} which can be associated with the quantum group $U_{q}(A_{2}^{(2)})$%
. The particle spectrum consists in general of a three-component kink and
its bound states, called breathers. For $L=3,4,6$ the aforementioned special
cases are recovered with $L=3$ being distinguished by the absence of any
kink state \cite{Sm}.\medskip

\begin{center}
\begin{tabular}{|l|l|l|l|l|}
\hline\hline
$L$ & conformal model & $c$ & $\Delta _{1,2}^{-1}$ & algebra \\ \hline\hline
$3$ & Ising in magnetic field & $1/2$ & $16$ & $E_{8}$ \\ \hline\hline
$4$ & tricritical Ising & $7/10$ & $10$ & $E_{7}$ \\ \hline\hline
$6$ & tricritical three-state Potts & $6/7$ & $7$ & $E_{6}$ \\ \hline\hline
$\infty $ & free boson & $1$ & $4$ & $D_{4}$ \\ \hline\hline
\end{tabular}
\smallskip
\end{center}

{\small \noindent \textbf{Table 2.} Displayed are three special cases of
minimal conformal models together with the values of the associated central
charge and the anomalous scaling dimension of the perturbing operator. All
of them are distinguished by an underlying Lie algebraic structure. In case
of the unrestricted model, L=}$\infty ${\small , one recovers the conformal
theory associated with a free boson.}\medskip

The universal quantity (\ref{amp}) can be extracted from the exact S-matrix
theories by applying the thermodynamic Bethe ansatz \cite{YY,TBAZ,KM,F}
which allows the two-particle scattering amplitude to be related to the
infinite bulk vacuum energy: 
\begin{equation}
\left\langle \Theta \right\rangle =\frac{2\pi m^{2}}{\delta _{1}}=-4\pi m^{2}%
\mathcal{Q}^{\pm }(0)\mathcal{S}^{\pm }(0)^{2},\quad -i\frac{d}{d\beta }\ln
S(\beta )=-\sum_{n=1}^{\infty }\delta _{n}\,e^{-n|\beta |}\;.  \label{tbaf}
\end{equation}
Here $\left\langle \Theta \right\rangle $ denotes the infinite volume vacuum
expectation value of the trace of the energy momentum tensor of the
associated integrable quantum field theory, $m$ the physical mass of the
fundamental particle (which can be identified with the spectral gap (\ref{xi}%
)) and $S(\beta )$ its two-particle scattering amplitude depending on the
rapidity $\beta $ parametrizing the two-momentum $p^{\mu }=m(\cosh \beta
,\sinh \beta )$. Formula (\ref{tbaf}) is obtained when computing the
groundstate energy of the integrable quantum field theory defined through
the exact S-matrix on a cylinder of circumference $R$. In the UV limit, $%
R\rightarrow 0$, the infinite bulk vacuum energy can be extracted by
comparison against results from conformal perturbation theory \cite{TBAZ,KM}%
. Equation (\ref{tbaf}) has to be applied with caution as it must be
modified when logarithmic terms are present in the singular part of the free
energy. It is, however, believed to hold true for the scaling field theories
and S-matrices mentioned and the amplitude (\ref{tbaf}) has been reported in 
\cite{F} based on the exact scattering matrix theory in \cite{Sm}. In
addition, from formula (\ref{tbaf}) it can be seen that the universal
amplitude enters the vacuum expectation value of the energy momentum tensor.
The latter is needed as important input information in the form factor
program \cite{FF,FFa}. Setting up a set of recursive functional equations it
allows all multi-particle form factors of the quantum field operators to be
constructed starting from their vacuum expectation values.

Our discussion of the dilute $A_{L}$ models provides an alternative and
independent way of calculating the universal amplitude (\ref{amp}) and is
consistent with formula (\ref{tbaf}) involving the scattering matrices \cite
{Sm} and their thermodynamic Bethe ansatz analysis. It is one of the main
results of our paper. \smallskip

We also reveal the close link between the exact S-matrix theory and the
transfer matrix method on another level which is of practical importance in
the calculation of the excitation spectrum and the spectral gaps (\ref{xi}).
As in the bootstrap approach we proceed by constructing functional relations
using entirely analytic arguments. In fact, we establish a one-to-one match
between the functional relations satisfied by the excitation spectrum of the
transfer matrix given in terms of elliptic functions and the scattering
amplitudes given in terms of trigonometric functions. Besides crossing
symmetry and unitarity, the crucial bootstrap identity characterising the
S-matrices \cite{Sm} is the functional relation associated with a
self-fusing process $b\times b\rightarrow b$: 
\begin{equation}
S_{bb}(\beta )S_{bb}(\beta +i\tfrac{2\pi }{3})=S_{bb}(\beta +i\tfrac{\pi }{3}%
)\;.  \label{Sboot}
\end{equation}
Here $S_{bb}$ denotes the two-particle amplitude either of the fundamental
particle or of a bound state of it. This distinction is important as, except
for the case $L=3$, the fundamental particle does not necessarily display a
self-fusing process. However, there are always bound states of the
fundamental particle which exhibit this peculiar property \cite{Sm}. The
equation (\ref{Sboot}) is understood w.r.t. analytic continuation and
implies a pole in the physical sheet at $\beta =i\frac{2\pi }{3}$.\smallskip

In the context of the dilute $A_{L}$ model we demonstrate in the
thermodynamic limit that for regime 2 the bootstrap equation (\ref{Sboot})
is reflected in the excitation spectrum of the dilute $A_{L}$ transfer
matrix by the appearance of an analogous identity, 
\begin{equation}
r(u)r(u+2\lambda )=r(u+\lambda )\;,\quad r(u)=\lim_{N\rightarrow \infty }%
\frac{\Lambda (u)}{\Lambda _{0}(u)}\;.  \label{rboot}
\end{equation}
The variable $u$ is the spectral parameter entering the Boltzmann weights
and does not have the physical interpretation of the rapidity in (\ref{Sboot}%
). The constant $\lambda $ is the crossing parameter whose dependence on the
maximal height value varies in the different branches of the model (see
Table 1). We stress that the relation (\ref{rboot}) holds for any excited
state $\Lambda $ not only the leading one. It has been reported before in
the literature for $L=3,4,6$ as an outcome of the exact perturbation theory
approach \cite{BS2,BS3,SB,SB3} based on various string hypotheses \cite
{BNW,GN,GNa,GNp}.

In this paper we derive it in the massive regime for general $L$ directly
from the known eigenvalue spectrum using only analytic arguments. Functional
equations of the transfer matrix in the form of an inversion relation are
well-known in the literature, e.g. \cite{Bx,P,KlZi,WPSN}. However, we
emphasize that the above functional identity (\ref{rboot}) of bootstrap type
is more powerful as it implies the inversion relation: 
\begin{equation}
r(u)r(u+3\lambda )=1\;.  \label{rinv}
\end{equation}
This relation corresponds to a combination of the quite general properties
of unitarity and crossing symmetry in exact S-matrix theory, while the
appearance of a bootstrap equation such as (\ref{rboot}) displays a
characteristic feature of the model. For $L=3,4,6,\infty $ we indeed recover
all bootstrap identities by employing a generic formulation in terms of
Coxeter geometry, originally applied in the context of affine Toda field
theories \cite{PD,FLO,FO}. The fundamental excitations of the dilute $%
A_{3,4,6,\infty }$ models can be cast into a single formula involving the
Coxeter element of the underlying Lie algebra, just as the $ADE$ affine Toda
S-matrices for real coupling have been \cite{FO,PD}. The latter contain the
aforementioned scattering matrices as special cases, with the important
difference of a coupling dependent CDD factor \cite{AFZ,MC,BCDS,DV,M}%
.\smallskip 

As in the context of the bootstrap construction of scattering matrices, the
functional equations describing the excitation spectrum determine it
completely, up to the location of its zeroes and poles. The missing
information on their position is encoded in the Bethe roots. We will argue
that in the thermodynamic limit the leading excitations are related either
to a hole or a two-string in the groundstate distribution of the Bethe
roots. This picture is supported by revealing the underlying Coxeter
geometry related to the exceptional algebras $E_{8,7,6}$ inherent to the
special cases $L=3,4,6$ and to the simple Lie algebra $D_{4}$ for the
unrestricted model $L=\infty $ (see Table 2). The latter is related to the
sine-Gordon model at special values of the coupling constant \cite{Sm}.
Moreover, our findings are consistent with all previous numerical
investigations in the literature \cite{BNW,GN,MO,BS} and the outcome of the
exact perturbation theory approach for $L=3,4,6$ \cite{BS2,BS3,SB,SB3}%
.\smallskip 

This article is organized as follows. In Section 2 we briefly review the
definition and the finite size eigenvalue spectrum of the dilute $A_{L}$
model. We also report a series expansion of the free energy density in the
elliptic nome based on earlier calculations of the groundstate \cite
{WPSN,BS2}. Section 3 states our analyticity assumptions and presents the
derivation of the bootstrap equation (\ref{rboot}) determining the explicit
form of the excitation spectrum in the thermodynamic limit up to its poles
and zeroes. While the discussion in Section 3 applies to any excited state,
Section 4 presents a concrete proposal for the Bethe root distributions of
the leading excitation in regime 2 in the form of a hole and a two-string.
The position of the zeroes is determined for these excitations and we
calculate the spectral gaps in the eigenspectrum as well as the associated
universal amplitude ratios. Section 5 compares the outcome against earlier
results for $L=3,4,6$ and presents new results on the unrestricted model $%
L=\infty $. The underlying Lie algebraic structures $E_{8,7,6}$ and $D_{4}$
are revealed by using Coxeter geometry which allows the fundamental
excitations of all four models to be expressed in a single formula. We also
derive numerous new identities which are in one-to-one correspondence with
the bootstrap identities of affine Toda field theory. The underlying Lie
algebraic structures will also be revealed in the derivation of the spectral
gaps associated with all $8,7,6,4$ fundamental particles. We obtain a series
expansion in terms of the ordering field (the elliptic nome) yielding the
corrections to the scaling behavior (\ref{xi}) in the off-critical regime.
The powers in the expansion match the affine Lie algebra exponents of $%
E_{8,7,6}$ and $D_{4}$. Section 6 states our conclusions.

\section{The off-critical dilute $A_{L}$ model}

The dilute $A_{L}$ model \cite{WNS} is an exactly solvable RSOS model
defined on the square lattice. Each site is allowed to take one of $L$
possible height values and the associated adjacency matrix is given by $3-A$
with $A$ being the Cartan matrix of the Lie algebra $A_{L}\equiv su(L+1)$.
In other words, the Boltzmann weights of a face are non-zero only if
neighbouring heights coincide or differ by $\pm 1$. Away from criticality
the Boltzmann weights are parametrised in terms of Jacobi's theta functions
of nome $p=e^{-\tau }$. We will frequently make use of their conjugate
modulus representation, 
\begin{equation}
\vartheta _{1}(u,e^{-\tau })=\sqrt{\frac{\pi }{\tau }}\,e^{-\left( u-\frac{%
\pi }{2}\right) ^{2}/\tau }E\left( e^{-2\pi u/\tau },e^{-2\pi ^{2}/\tau
}\right)  \label{theta1}
\end{equation}
with 
\begin{equation}
E(z,p)=\prod_{n=1}^{\infty }(1-p^{n-1}z)(1-p^{n}/z)(1-p^{n})\;.  \label{E}
\end{equation}
The explicit expression of the Boltzmann weights will not matter in our
discussion and can be found in \cite{WPSN}. The eigenspectrum of the
row-to-row transfer matrix for a strip of length $N$ has been reported in 
\cite{BNW} (we choose $N$ even), 
\begin{eqnarray}
\Lambda (u) &=&\omega \left\{ \frac{\vartheta _{1}(2\lambda -u,p)\vartheta
_{1}(3\lambda -u,p)}{\vartheta _{1}(2\lambda ,p)\vartheta _{1}(3\lambda ,p)}%
\right\} ^{N}G(u)  \notag \\
&&+\left\{ \frac{\vartheta _{1}(u,p)\vartheta _{1}(3\lambda -u,p)}{\vartheta
_{1}(2\lambda ,p)\vartheta _{1}(3\lambda ,p)}\right\} ^{N}\frac{G(u-\lambda )%
}{G(u-2\lambda )}  \notag \\
&&+\omega ^{-1}\left\{ \frac{\vartheta _{1}(u,p)\vartheta _{1}(\lambda -u,p)%
}{\vartheta _{1}(2\lambda ,p)\vartheta _{1}(3\lambda ,p)}\right\}
^{N}G(u-3\lambda )^{-1}  \label{eig}
\end{eqnarray}
Here we have introduced for convenience the auxiliary function 
\begin{equation}
G(u)=\prod_{n=1}^{N}\frac{\vartheta _{1}(u-u_{n}+\lambda ,p)}{\vartheta
_{1}(u-u_{n}-\lambda ,p)}\;.  \label{G}
\end{equation}
The behaviour of the model depends crucially on the choice of parameters.
Henceforth we will restrict ourselves to the regime 2 with the interval of
the spectral parameter and the crossing parameter $\lambda $ as specified in
Table 1. The complex numbers $u_{n}\,,\;n=1,...,N$ characterising the
eigenvalue are the Bethe roots, subject to the Bethe equations 
\begin{equation}
\frak{p}(u_{m})=-\omega \quad \text{with\quad }\frak{p}(u)=\left\{ \frac{%
\vartheta _{1}(u+\lambda ,p)}{\vartheta _{1}(u-\lambda ,p)}\right\} ^{N}%
\frac{G(u)}{G(u+\lambda )G(u-\lambda )} .  \label{p}
\end{equation}
where the phase factor $\omega =\exp \left( i\pi \frac{\ell }{L+1}\right) $
with $\ell =1,\ldots ,L$ depends on the particular position of the Bethe
roots chosen. In fact, using the known transformation properties of the
theta function 
\begin{eqnarray}
\vartheta _{1}\left( u\pm i\tau ,e^{-\tau }\right) &=&-e^{\pm \frac{\tau }{2}%
}e^{\mp 2iu}\vartheta _{1}\left( u,e^{-\tau }\right) \;,\quad \tau >0
\label{T1} \\
\vartheta _{1}\left( u+\pi ,p\right) &=&-\vartheta _{1}\left( u,p\right)
=\vartheta _{1}\left( -u,p\right)  \label{T2}
\end{eqnarray}
one verifies that the corresponding eigenvalue $\Lambda (u)$ is invariant
under the simultaneous replacement 
\begin{equation}
u_{n}\rightarrow u_{n}\pm i\tau \quad \text{and}\quad \omega \rightarrow
\omega \,e^{\mp i4\lambda }  \label{t1}
\end{equation}
as well as under the shift 
\begin{equation}
u_{n}\rightarrow u_{n}\pm \pi  \label{t2}
\end{equation}
of a single Bethe root. Moreover, one easily deduces from the transformation
properties 
\begin{equation}
G(u)=e^{\pm 4Ni\lambda }G(u\pm i\tau )=G(u+\pi )  \label{TG}
\end{equation}
of the auxiliary function (\ref{G}) that the eigenvalues satisfy 
\begin{equation}
\Lambda (u+\pi )=\Lambda (u)\quad \text{and}\quad \Lambda (u\pm i\tau
)=p^{\mp N}e^{\pm i6\lambda N}e^{\mp 4Niu}\Lambda (u)  \label{period1}
\end{equation}
while the function determining the Bethe equations (\ref{p}) is doubly
periodic 
\begin{equation}
\frak{p}(u)=\frak{p}(u+i\tau )=\frak{p}(u+\pi )\;.
\end{equation}
Finally, from the identity $\vartheta _{1}\left( u,e^{i\pi }p\right) =e^{i%
\frac{\pi }{4}}\vartheta _{1}\left( u,p\right) $ it is seen that, for all
finite $N$, the eigenspectrum and Bethe equations are invariant under the
exchange $p\rightarrow e^{i\pi }p$ as only ratios of Jacobi's theta
functions occur in (\ref{eig}) and (\ref{p}). However, the Boltzmann weights 
\cite{WNS} are in general not invariant under sign reversal of the elliptic
nome, but can be related through a relabelling of the heights if $L$ is odd.
Therefore, while all eigenvalues of the transfer matrix can be parametrized
through (\ref{eig}) and (\ref{p}) with $p$ replaced by $|p|$, an eigenvalue
present for $p>0$ might be absent for $p<0$, and vice versa, if $L$ is even.

\subsection{The groundstate and scaling corrections}

The eigenvalue $\Lambda _{0}$ corresponding to the groundstate of the model
has been investigated for arbitrary $L$ in \cite{BS2} by applying Baxter's
exact perturbation theory approach \cite{Baxter} in the ordered limit $%
|p|\rightarrow 1$. The groundstate Bethe roots are assumed to lie on the
imaginary axis, which has been supported by numerical investigations. The
result obtained for the groundstate is assumed to hold for any $0<|p|<1$ in
the thermodynamic limit and allows one to compute the free energy density (%
\ref{free}) 
\begin{eqnarray}
f &=&-\lim_{N\rightarrow \infty }N^{-1}\ln \Lambda _{0}(\tfrac{3\lambda }{2}%
)=4\sqrt{3}\sum_{k=0}^{\infty }\frac{\cos \left[ (\frac{5\pi }{6}-\frac{\pi
^{2}}{6\lambda })(6k+1)\right] }{(6k+1)\sin \left[ \frac{\pi ^{2}}{6\lambda }%
\,(6k+1)\right] }\,\frac{p^{\frac{\pi }{3\lambda }(6k+1)}}{1-p^{\frac{\pi }{%
3\lambda }(6k+1)}}  \notag \\
&&-4\sqrt{3}\sum_{k=1}^{\infty }\frac{\cos \left[ (\frac{5\pi }{6}-\frac{\pi
^{2}}{6\lambda })(6k-1)\right] }{(6k-1)\sin \left[ \frac{\pi ^{2}}{6\lambda }%
\,(6k-1)\right] }\,\frac{p^{\frac{\pi }{3\lambda }(6k-1)}}{1-p^{\frac{\pi }{%
3\lambda }(6k-1)}}  \notag \\
&&-8\sum_{k=1}^{\infty }\frac{\sin ^{2}\frac{3\lambda }{2}k}{k}\frac{\cos
5\lambda k\cos \lambda k}{\cos 3\lambda k}\,\frac{p^{2k}}{1-p^{2k}}%
=f_{s}+f_{ns}.  \label{f}
\end{eqnarray}
This formula is obtained from the result for the partition function per site
derived in \cite{WPSN,BS2} by applying the Poisson summation formula and the
residue theorem. The spectral parameter has been set to the isotropic value $%
u=3\lambda /2$ in order to ensure that the eigenvalues are real. We come
back to this point later on. From the series expansion (\ref{f}) the
coefficient of the leading term in the critical limit $p\rightarrow 0^{+}$
is obtained as 
\begin{equation}
\mathcal{Q}^{+}(0)=\lim_{p\rightarrow 0^{+}}p^{-\frac{1}{1-\Delta }}f_{s}=\,4%
\sqrt{3}\,\frac{\cos \frac{\pi }{6\lambda }(5\lambda -\pi )}{\sin \frac{\pi 
}{6\lambda }\pi }  \label{f0}
\end{equation}
with $2\Delta =2\Delta _{1,2}=2-6\lambda /\pi $ being the scaling dimension
of the perturbing operator. As already pointed out in the introduction, the
latter depends on the chosen normalization of the fields and is a
non-universal quantity. In order to obtain the universal quantity (\ref{amp}%
) we need to compute (\ref{xi}) from the leading excitation in the
thermodynamic limit. To this end, in the next section we derive functional
equations satisfied by any excited state in the thermodynamic limit.

\section{Bootstrap identities in the dilute $A_{L}$ model}

Conventionally the excitation spectrum in the thermodynamic limit is
calculated proceeding via the thermodynamic Bethe ansatz. Alternatively, one
might proceed entirely by analytic arguments as has been demonstrated e.g.
in the context of the XYZ model \cite{KlZi}. Rather than solving the
intricate non-linear integral equations of the thermodynamic Bethe ansatz,
we will instead derive a set of functional relations for the ratio 
\begin{equation}
r(u)=\lim_{N\rightarrow \infty }\frac{\Lambda (u)}{\Lambda _{0}(u)}\;.
\label{rdef}
\end{equation}
Here $\Lambda (u)$ denotes the eigenvalue of any excited state over the
ground state $\Lambda _{0}(u)$ such that the ratio $\Lambda /\Lambda _{0}$
has a non-vanishing well-defined limit as $N\rightarrow \infty $. We will in
particular show the functional equation (\ref{rboot}), which, in conjunction
with certain analyticity assumptions, imposes powerful restrictions
determining the excitation spectrum up to the position of its poles and
zeroes. The latter are fixed by the Bethe roots which are usually unknown.
However, the functional relations we are going to discuss yield constraints
on their positions.

\subsection{Asymptotic behaviour in the thermodynamic limit}

Recall that the Bethe roots $\{u_{n}^{(0)}\}$ belonging to the groundstate $%
\Lambda _{0}$ lie on the imaginary axis. It is reasonable to assume in
general that for the low lying excited states $\Lambda $ the associated
Bethe root distribution $\{u_{n}\}$ is similar, and that at most a finite
number of holes are present or a finite number of Bethe roots have
non-vanishing real part. Well away from the critical point, $0<|\ln p|\ll 1$%
, this information is sufficient to show that as $N\rightarrow \infty $
either the first, second or third term in the eigenvalue expression (\ref
{eig}) becomes dominant as the spectral parameter sweeps through the allowed
interval $0<u<3\lambda $. The crucial feature exploited is the conjugate
modulus transformation (\ref{theta1}), as outlined in Appendix A.
Explicitly, one finds for the groundstate as well as for the excited states
the following: 
\begin{equation}
\Lambda (u)\sim \left\{ 
\begin{array}{cc}
\omega \left\{ \frac{\vartheta _{1}(2\lambda -u)\vartheta _{1}(3\lambda -u)}{%
\vartheta _{1}(2\lambda )\vartheta _{1}(3\lambda )}\right\} ^{N}G(u)\left[
1+o(e^{-N})\right] , & 0<u<\lambda \\ 
\left\{ \frac{\vartheta _{1}(u)\vartheta _{1}(3\lambda -u)}{\vartheta
_{1}(2\lambda )\vartheta _{1}(3\lambda )}\right\} ^{N}\frac{G(u-\lambda )}{%
G(u-2\lambda )}\left[ 1+o(e^{-N})\right] , & \lambda <u<2\lambda \\ 
\frac{1}{\omega }\left\{ \frac{\vartheta _{1}(u)\vartheta _{1}(\lambda -u)}{%
\vartheta _{1}(2\lambda )\vartheta _{1}(3\lambda )}\right\} ^{N}\frac{1}{%
G(u-3\lambda )}\left[ 1+o(e^{-N})\right] , & 2\lambda <u<3\lambda
\end{array}
\right. \,.  \label{asym}
\end{equation}
By the above notation we indicate that all terms besides the leading one are
exponentially suppressed in $N$. In order to discuss the thermodynamic limit
we introduce 
\begin{equation}
g(u):=\lim_{N\rightarrow \infty }g^{N}(u)\;,\quad g^{N}(u):=\frac{G(u)}{%
G_{0}(u)}\;,  \label{g}
\end{equation}
where $G_{0}(u)$ denotes the auxiliary function (\ref{G}) corresponding to
the groundstate. By an analogous argument as in the XYZ case \cite{KlZi},
the convergence of the above limit is expected to be uniform as only a
finite number of Bethe roots have non-vanishing real part or are missing
compared to the groundstate distribution. The excitation spectrum then obeys 
\begin{equation}
r(u)=\lim_{N\rightarrow \infty }\frac{\Lambda (u)}{\Lambda _{0}(u)}=\left\{ 
\begin{array}{cc}
\frac{\omega }{\omega _{0}}\,g(u)\;, & 0<\func{Re}u<\lambda \\ 
g(u-\lambda )/g(u-2\lambda )\;, & \lambda <\func{Re}u<2\lambda \\ 
\frac{\omega _{0}}{\omega }\,g(u-3\lambda )^{-1}\;, & 2\lambda <\func{Re}%
u<3\lambda
\end{array}
\right. .  \label{B}
\end{equation}
An obvious consequence of this limiting behaviour is the identity 
\begin{equation}
r(u)r(u+2\lambda )=r(u+\lambda )\;,\quad 0<\func{Re}u<\lambda .
\label{Boot0}
\end{equation}
In order to promote this relation to the functional equation (\ref{rboot})
valid in the whole complex plane, we need to investigate the analytic
behaviour of the excitation spectrum in the thermodynamic limit.

\subsection{Analyticity and the general form of the excitation spectrum}

From the explicit expression (\ref{eig}) for the eigenspectrum, it is
inferred that, for finite $N,$ poles and zeroes of the ratio $\Lambda
(u)/\Lambda _{0}(u)\;$are completely determined by the location of the Bethe
roots. Employing the asymptotic behaviour (\ref{asym}) one easily deduces
that for $N\gg 1$ the purely imaginary Bethe roots $u_{n}$ of the excited
and of the ground state give rise to poles located at 
\begin{equation}
u_{\func{Im}}^{P}=u_{n}+\lambda ,\;u_{n}+2\lambda \,\func{mod}\pi \;.
\end{equation}
These pole contributions will cancel each other when the limit $N\rightarrow
\infty $ is taken, since the imaginary roots then become densely located
along the imaginary axis centered at $\lambda ,2\lambda \func{mod}\pi $. The
only exceptions are poles in the groundstate which give rise to zeroes of
the eigenvalue ratio $\Lambda (u)/\Lambda _{0}(u)$ when holes are present in
the excited state. Bethe roots $u_{n}$ of the excited state belonging to a
string, i.e. $\func{Re}u_{n}\neq 0$, will also give rise to isolated zeroes
and poles. However, since the number of holes and strings is assumed to be
finite, one infers from (\ref{eig}) that in each of the intervals 
\begin{equation}
0<\func{Re}u<\lambda \;,\quad \lambda <\func{Re}u<2\lambda \quad \text{%
and\quad }2\lambda <\func{Re}u<3\lambda
\end{equation}
the ratio $\Lambda (u)/\Lambda _{0}(u)$ is analytic except for isolated
points, and therefore converges towards a meromorphic function $%
r(u)=\lim_{N\rightarrow \infty }\Lambda (u)/\Lambda _{0}(u)$ in these
intervals. Upon analytic continuation $r(u)$ can then be extended to the
whole complex plane and the identity (\ref{Boot0}) becomes the functional
equation (\ref{rboot}) valid for all complex values of the spectral
parameter. As an immediate result we have the inversion relation (\ref{rinv}%
). Applying the inversion relation twice we deduce that the excitation
spectrum is doubly periodic: 
\begin{equation}
r(u)=r(u+i\tau )=r(u+6\lambda )\;.  \label{period}
\end{equation}
The first period is a direct consequence of (\ref{period1}) and is already
present at finite $N$. The assumption of analyticity together with the
double-periodicity property are powerful restrictions which determine the
excitation spectrum up to the position of its poles and zeroes by applying
the following theorem \cite{Baxter}:\medskip

\noindent \textbf{Theorem}. \emph{Let }$f(u)$\emph{\ be a meromorphic
function satisfying } 
\begin{equation*}
f(u+2K)=f(u+2iK^{\prime })=f(u)
\end{equation*}
\emph{with }$K,K^{\prime }$\emph{\ being quarter periods to the elliptic
modulus }$k$\emph{. If }$f(u)$\emph{\ has }$n$\emph{\ poles }$u_{j}^{P}$%
\emph{\ per period rectangle then it has also }$n$\emph{\ zeroes }$u_{j}^{Z}$%
\emph{\ and can be expressed as } 
\begin{equation*}
f(u)=Ce^{\frac{i\pi s^{\prime }}{K}u}\prod_{j=1}^{n}\frac{\vartheta _{1}%
\left[ \frac{\pi }{2K}(u-u_{j}^{Z}),e^{-\pi \frac{K^{\prime }}{K}}\right] }{%
\vartheta _{1}\left[ \frac{\pi }{2K}(u-u_{j}^{P}),e^{-\pi \frac{K^{\prime }}{%
K}}\right] }\;.
\end{equation*}
\emph{Here }$C$\emph{\ is a constant and }$s^{\prime }$\emph{\ an integer
determined by the sum rule } 
\begin{equation}
\sum_{j=1}^{n}(u_{j}^{Z}-u_{j}^{P})=2sK-2is^{\prime }K^{\prime }\;,\quad
s,s^{\prime }\in \mathbb{Z\;}.
\end{equation}
\medskip

\noindent We therefore conclude that the excitation spectrum of the dilute $%
A_{L}$ model at infinite volume is of the form 
\begin{equation}
r(u)=Ce^{\frac{i\pi s^{\prime }}{3\lambda }u}\prod_{j=1}^{n}\frac{\vartheta
_{1}\left[ \frac{\pi }{6\lambda }(u-u_{j}^{Z}),|p|^{\frac{\pi }{6\lambda }}%
\right] }{\vartheta _{1}\left[ \frac{\pi }{6\lambda }(u-u_{j}^{P}),|p|^{%
\frac{\pi }{6\lambda }}\right] }\;.  \label{Ex}
\end{equation}
Inserting this expression in the functional relations (\ref{rinv}) and (\ref
{rboot}) we deduce several constraints on the location of poles and zeroes.
From the inversion relation it follows that 
\begin{equation}
C^{2}=1,\;s^{\prime }=0,\;s=3\lambda n\quad \text{and\quad }%
u_{j}^{P}=u_{j}^{Z}\pm 3\lambda \;.
\end{equation}
It is therefore sufficient to locate either the zeroes or the poles. From
the functional equation (\ref{rboot}) we deduce that the existence of a zero 
$u_{j}^{Z}$ implies another zero either at $u_{j}^{Z}+\lambda $ or at $%
u_{j}^{Z}-\lambda $. Thus, all zeroes come in pairs and $n$ in (\ref{Ex}) is
even. Furthermore, the constant in (\ref{Ex}) is fixed to be $C=(-1)^{n/2}$.

\subsection{Hermitian analyticity and unitarity}

In order to restrict the position of zeroes and poles further we derive
another functional equation which is the analog of hermitian analyticity and
unitarity in exact S-matrix theory \cite{herman}: 
\begin{equation}
r(-u^{\ast })^{\ast }=r(u)^{-1}\;.  \label{HA}
\end{equation}
Here $\ast $ denotes complex conjugation. In order to prove (\ref{HA}) we
need to make the further assumption that the complex Bethe roots form a set
which is symmetric w.r.t. to the imaginary axis modulo multiples of $\pi $,
i.e. the set $\{u_{n}\}$ should be invariant under the mapping 
\begin{equation}
u_{n}\rightarrow -u_{n}^{\ast }\;\func{mod}\pi \;.  \label{s1}
\end{equation}
This is trivially fullfilled for the Bethe roots of the groundstate $\Lambda
_{0}$, which lie on the imaginary axis. In the case of an excited state $%
\Lambda $ a finite number of the Bethe roots $\{u_{n}\}$ might have
non-vanishing real part and the above constraint is non-trivial. However
there is numerical evidence for $L=3,4$ that (\ref{s1}) is satisfied and all
string hypotheses \cite{BNW,GN,GNp,SB3} previously reported in the
literature for $L=3,4$ are subject to this constraint. As a consequence of (%
\ref{s1}) the auxiliary function (\ref{G}) for the groundstate as well as
for an excited state satisfies 
\begin{equation}
G(-u)^{\ast }=\prod_{n=1}^{N}\frac{\vartheta _{1}(u^{\ast }+u_{n}^{\ast
}-\lambda )}{\vartheta _{1}(u^{\ast }+u_{n}^{\ast }+\lambda )}=G(u^{\ast
})^{-1}\;.  \label{haG}
\end{equation}
From this property we infer for all $N$ that the corresponding eigenvalues (%
\ref{eig}) are crossing symmetric: 
\begin{equation}
\Lambda (u)=\Lambda (3\lambda -u^{\ast })^{\ast }\;.  \label{cross}
\end{equation}
Upon analytic continuation in the thermodynamic limit we conclude from this
crossing symmetry and the inversion relation (\ref{rinv}) that the property (%
\ref{HA}) must hold. Thus, we can eliminate the pole variables $u^{P}$ from (%
\ref{Ex}) by setting $u^{P}=-u^{Z\,\ast }$. We further deduce from crossing
symmetry that the set of zeroes of the excitation spectrum (\ref{Ex}) must
be invariant under the mapping 
\begin{equation}
u^{Z}\rightarrow 3\lambda -u^{Z\,\ast }\;.  \label{crossZ}
\end{equation}
Together with our earlier observation that the zeroes come at least in pairs 
$(u^{Z},u^{Z}+\lambda )$ this now implies that one has either a pair 
\begin{equation}
(u^{Z},u^{Z}+\lambda )\quad \text{with\quad }\func{Re}u^{Z}=\lambda
\label{Z1}
\end{equation}
or a quartet of zeroes 
\begin{equation}
(u^{Z},u^{Z}+\lambda ,2\lambda -u^{Z\,\ast },3\lambda -u^{Z\,\ast })\;.
\label{Z2}
\end{equation}
As argued earlier, the real parts of the zeroes $u^{Z}$ are determined by
the complex Bethe roots $u_{n}$ with $\func{Re}u_{n}\neq 0$, whence the
conditions (\ref{Z1}) and (\ref{Z2}) constitute further restrictions on the
string structure of the dilute $A_{L}$ model, besides (\ref{s1}). We leave a
complete investigation of the string structure to future work \cite{KSP} and
now focus on the leading excitations in regime 2.

\section{The leading excitations in regime 2}

Having determined the general form of the excitation spectrum in regime 2 in
the thermodynamic limit, we now proceed by putting forward a concrete
proposal for the leading excitations which will ultimately allow us to
compute the leading correlation length (\ref{xi}) and the universal quantity
(\ref{amp}).\medskip 

\noindent \textbf{Proposal. }\emph{The leading excitation in regime 2
consists either of a hole or the presence of a two-string in the groundstate
distribution of the Bethe roots.}\medskip

This proposal is supported by numerical calculations \cite{BNW,GN,MO,BS},
the thermodynamic Bethe ansatz \cite{BNW} and the exact perturbation theory
approach for the dilute $A_{L}$ models in regime 2 with $L=3,4$ \cite
{BS2,BS3,SB,SB3}. Further support comes from extracting the correct Lie
algebraic structures for $L=3,4,6,\infty ,$ as we will discuss in the
subsequent section. Here we first discuss the implications of our proposal
for general $L$.\medskip 

From (\ref{asym}) and (\ref{B}) we infer that the zeroes of the excitation
spectrum are determined by the zeroes of the auxiliary function (\ref{g}) in
the appropriate domain of analyticity. The latter are determined by the
Bethe roots according to the definition (\ref{G}). In the thermodynamic
limit only the real part of the zeroes $u^{Z}$ in the excitation spectrum (%
\ref{Ex}) needs to be determined, since $\func{Im}u^{Z}$ is linked to the
imaginary part of the Bethe roots which can be parametrized by a continuous
variable as $N\rightarrow \infty $. Given a Bethe root we note that $G(u)$
in general has a simple zero and pole at 
\begin{equation}
G(u):\quad u_{I}^{Z}=u_{n}-\lambda \quad \text{and\quad }u_{I}^{P}=u_{n}+%
\lambda ,  \label{G1}
\end{equation}
respectively. Accordingly, the second and third terms in the eigenvalue
expression (\ref{eig}) exhibit zeroes and poles located at 
\begin{equation}
\frac{G(u-\lambda )}{G(u-2\lambda )}:\quad u_{II}^{Z}=u_{n},u_{n}+3\lambda
\;,\quad u_{II}^{P}=u_{n}+\lambda ,u_{n}+2\lambda  \label{G2}
\end{equation}
and 
\begin{equation}
G(u-3\lambda )^{-1}:\quad u_{III}^{Z}=u_{n}+4\lambda \;,\text{\quad }%
u_{III}^{P}=u_{n}+2\lambda \;.  \label{G3}
\end{equation}
We deduce from the asymptotic behaviour (\ref{asym}) that in the
thermodynamic limit these poles and zeroes are only of significance for the
excitation spectrum if their real parts are located inside the domains of
analyticity stated in (\ref{B}). This constraint originates in the fact that
the excitation spectrum (\ref{Ex}) is the analytic continuation of
meromorphic functions whose region of analyticity is restricted. The
resulting zeroes are therefore characteristic for the regime chosen. Recall
from (\ref{TG}) that the zeroes and poles of the auxiliary function are only
fixed modulo $\pi .$ The following relation for regime 2 will therefore
prove important to determine the zeroes and poles of the excitation
spectrum: 
\begin{equation}
\text{regime 2}:\quad \pi =3\lambda +\frac{\pi }{r}\;,\quad r=\Delta
_{1,2}^{-1}=4\frac{L+1}{L-2}.
\end{equation}
Henceforth, it will always be understood that zeroes related by multiples of 
$\pi $ can be identified.

\subsubsection{Zeroes from imaginary Bethe roots and holes}

Before we turn to complex Bethe roots with nonvanishing real part, we first
investigate the possible zeroes and poles related to the imaginary roots in
the thermodynamic limit. As we have established in (\ref{G1}), (\ref{G2})
and (\ref{G3}), only the poles $u_{n}+\lambda ,u_{n}+2\lambda $ may occur as 
$N\rightarrow \infty $. Note that these poles are located at the border of
the domains of analyticity but are common to both relevant terms in (\ref
{asym}). These poles cancel against the ones from the groundstate function
unless there is a hole in the excited state in the thermodynamic limit. Then
we obtain zeroes with real part located at 
\begin{equation}
\func{Re}u^{Z}=\lambda ,2\lambda \;  \label{Zh}
\end{equation}
in accordance with our earlier observation (\ref{Z1}). Whether or not a hole
is present can in principle be inferred from the Bethe equations, which we
will discuss below in the ordered limit. In general a hole might be
accompanied by a string, i.e. some of the Bethe roots might in addition have
vanishing real part. This picture has been suggested by unpublished
numerical calculations \cite{GNp} for $L=4$ where the presence of a hole
excitation has been reported in connection with the following 3-string: 
\begin{equation}
v,v\pm 2\lambda \;,\quad \func{Re}v=\tfrac{\pi}{2} .  \label{3s}
\end{equation}
This scenario has been investigated by the exact perturbation theory
approach \cite{SB3} and it was observed that the three string does not of
itself contain any relevant information. We confirm this outcome in our
analytic approach; from (\ref{G1}), (\ref{G2}), and (\ref{G3}) we deduce for
general $L$ that the proposed 3-string does not give rise to any zeroes in
the excitation spectrum.

Further restrictions on the existence of a hole excitation might originate
in the maximal height value $L$ and in the choice $p>0$ or $p<0$. In fact,
when we evaluate these zeroes for $L=3,4,6,\infty $ in a subsequent section
we will see that it does not give the leading excitation for $L=3$. This is
consistent with the suggested picture from scattering theory \cite{Sm} if we
associate the hole-excitation with a kink state, absent for the Ising model
in a magnetic field. The fundamental particle in the case when the hole
excitation is not present should be associated with a particular two-string,
which we consider next.

\subsubsection{Zeroes induced by a two-string}

A general two-string subject to the constraint (\ref{s1}) is of the form 
\begin{equation}
v\pm x,\quad \func{Re}v=0\,\text{or\thinspace }\tfrac{\pi }{2}\,,\;\func{Im}%
x=0\;.
\end{equation}
Setting $x=\lambda /2$ and $\func{Re}v=\pi /2$ this type of string has been
observed numerically for $L=3,4$ \cite{BNW,GN,GNp}. We propose it here for
general $L$ and show below that it leads to the correct algebraic structure
for the special cases $L=3,4,6,\infty $. In comparison with the exact
S-matrix theory proposed in \cite{Sm}, the associated excited state of the
dilute $A_{L}$ model should be identified with the lowest breather, i.e. a
bound state of two kinks.

One easily verifies that $G(u)$ has simple zeroes and poles at 
\begin{equation}
u_{I}^{Z}=v-\tfrac{3\lambda }{2},v-\tfrac{\lambda }{2}\quad \text{and\quad }%
u_{I}^{P}=v+\tfrac{\lambda }{2},v+\tfrac{3\lambda }{2}\;.
\end{equation}
Since $\func{Re}v=\pi /2$ the second zero has real part greater than $%
\lambda $ and thus does not constitute a zero of the eigenvalue in the
thermodynamic limit. Note also that the poles are not present in the
thermodynamic limit. The second term and third terms in the eigenvalue
expression (\ref{eig}) exhibit zeroes and poles at 
\begin{eqnarray}
u_{II}^{Z} &=&v-\tfrac{\lambda }{2},v+\tfrac{\lambda }{2},v-\tfrac{\lambda }{%
2}-\tfrac{\pi }{r},v+\tfrac{\lambda }{2}-\tfrac{\pi }{r} \\
u_{II}^{P} &=&v+\tfrac{\lambda }{2},\left( v+\tfrac{3\lambda }{2}\right)
^{2},v-\tfrac{\lambda }{2}-\tfrac{\pi }{r}  \notag
\end{eqnarray}
and 
\begin{equation}
u_{III}^{Z}=v+\tfrac{\lambda }{2}-\tfrac{\pi }{r},v+\tfrac{3\lambda }{2}-%
\tfrac{\pi }{r}\quad \text{and\quad }u_{III}^{P}=v+\tfrac{3\lambda }{2},v+%
\tfrac{5\lambda }{2}\;.
\end{equation}
We therefore deduce the following location of the real parts of the zeroes
in the excitation spectrum when the thermodynamic limit is taken: 
\begin{equation}
\func{Re}u^{Z}=\tfrac{\pi }{2r},\lambda +\tfrac{\pi }{2r},2\lambda -\tfrac{%
\pi }{2r},3\lambda -\tfrac{\pi }{2r}\;.  \label{Zs}
\end{equation}
Note that the resulting quartet fulfills the condition (\ref{Z2}).

Having determined the zeroes of the excitation spectrum we are now in the
position to write down the corresponding ratio of eigenvalues (\ref{Ex}) and
to calculate the resulting spectral gaps. Before we come to this point we
further support our proposal on the excitations and their Bethe root
structure by investigating the Bethe equations (\ref{p}) in the ordered
limit.

\subsection{Bethe equations in the ordered limit}

As solving the Bethe equations directly is a complicated matter, we consider
them in the ordered limit $|p|\rightarrow 1$ where we can check the proposed
excitations for consistency by investigating the asymptotic behavior as $%
N\rightarrow \infty $. We will show that the function (\ref{p})
characterizing the Bethe equations is a pure phase in the massive regime for 
$N\gg 1$ as required. Its asymptotic behavior is determined by the real part
of the Bethe roots which is the essential information entering the
excitation spectrum (\ref{Ex}). The latter depends continuously on the
elliptic nome, whence we expect that our findings in the ordered limit give
a qualitative picture valid for all values of $|p|$.

As we did for the eigenspectrum of the transfer matrix in Appendix A, it is
also of advantage here to rewrite the Bethe equations using the conjugate
modulus transformation (\ref{E}): 
\begin{equation}
\frak{p}(w)=\left\{ -\frac{E(x^{2s}w)}{wE(x^{2s}/w)}\right\}
^{N}\prod_{n=1}^{N}w_{n}^{\frac{2s}{r}}\frac{E(x^{2s}w/w_{n})E(x^{4s}w_{n}/w)%
}{E(x^{2s}w_{n}/w)E(x^{4s}w/w_{n})}\;.  \label{BEcm}
\end{equation}
We recall the definition of the variables, $w=e^{-\frac{2\pi }{\tau }u},\;s=%
\frac{L+2}{L-2}$ and $x=e^{-\frac{\pi ^{2}}{\tau r}}$ with $r=4\frac{L+1}{L-2%
}$. The dependence on the elliptic nome $x^{2r}$ is suppressed in the
notation.

\subsubsection{Bethe equations and the occurrence of holes}

Let us start by investigating the ordered limit $x\rightarrow 0$ ($%
|p|\rightarrow 1$) when only purely imaginary Bethe roots are present. From
the conjugate modulus representation (\ref{BEcm}) we infer that the leading
contribution for any Bethe root $w_{m}=e^{-\frac{2\pi }{\tau }u_{m}}$ is
given by 
\begin{equation}
\frak{p}(w_{m})\sim (-w_{m})^{-N}\prod\limits_{n=1}^{N}w_{n}^{\frac{2s}{r}%
}=o(1)\;.  \label{ho1}
\end{equation}
Thus, $\frak{p}(w_{m})$ lies on the unit circle as expected. Since we are
only interested in the thermodynamic limit, the exact position of the Bethe
roots does not matter. But we can extract an additional piece of qualitative
information, namely that the $N$ Bethe roots $w_{m}$ are completely
determined by an equation of order $N$. This shows that holes are absent for
any finite $N$ when only imaginary roots are present. The situation changes
when we consider in addition Bethe roots which have non-vanishing real part,
since they alter the asymptotic behavior in the ordered limit.

We have mentioned before the possible occurrence of the 3-string (\ref{3s})
for $L=4$ \cite{GNp,SB3} which according to our analysis does not give rise
to any zeroes in the excitation spectrum for general $L$. Its significance
is the induction of a hole in the purely imaginary Bethe root distribution
as observed in \cite{SB3} for $L=4$. We briefly review the argument for
general $L$. Set $w_{N}=z\,x^{3s+1},w_{N-1}=z^{\prime
}x^{-s+1},w_{N-2}=z^{\prime \prime }x^{7s+1}$ with $z,z^{\prime },z^{\prime
\prime }$ being pure phases up to exponentially small corrections in $N$. As
the remaining Bethe roots $w_{n},n<N-2$ also lie on the unit circle, one
deduces from (\ref{BEcm}) the following leading term in the ordered limit ($%
N $ even $\gg 1$), 
\begin{equation}
\frak{p}(w_{m})\sim w_{m}^{-N+2}\frac{(zz^{\prime }z^{\prime \prime })^{%
\frac{2s}{r}}}{z^{2}}\prod\limits_{n=1}^{N-3}w_{n}^{\frac{2s}{r}%
}=o(1)\;,\quad 1\leq m<N-2  \label{ho2}
\end{equation}
As required $\frak{p}(w_{m})$ lies on the unit circle. In contrast to (\ref
{ho1}), however, the $N-3$ imaginary Bethe roots are now determined by an
equation of order $N-2$. The extra solution points towards the existence of
a hole. In addition to the Bethe equations (\ref{ho2}) one also needs to
verify that the equations for the complex Bethe roots $w_{N},w_{N-1},w_{N-2}$
are of order one in the ordered limit. The product of the Bethe equations
corresponding to the three roots in the string has to be taken in order to
avoid complications in the thermodynamic limit where terms either in the
numerator or denominator become small. Again one finds that the leading term
lies on the unit circle in the ordered limit, $\frak{p}(w_{N-2})\frak{p}%
(w_{N-1})\frak{p}(w_{N})\sim o(1)$, showing that the string structure is
consistent.

\subsubsection{Bethe equations for the two-string excitation}

We now consider the Bethe roots for the excitation associated with the
two-string. We choose the first $N-2$ Bethe roots $w_{n}$ to lie on the unit
circle and set $w_{N-1}=z\,x^{2s+1}$, $w_{N}=z^{\prime }x^{4s+1}$. In the
thermodynamic limit we have $|z|=|z^{\prime }|=1$ and $z=z^{\prime }$, thus
for sufficiently large $N$ we may assume these properties up to
exponentially small corrections in $N$. For the Bethe roots $w_{m},m<N-1$ we
find ($6s+2=2r$) 
\begin{equation}
\frak{p}(w_{m})\sim (-w_{m})^{2-N}z^{\frac{4s}{r}-2}\prod_{n=1}^{N-2}w_{n}^{%
\frac{2s}{r}}=o(1)\;,\quad x\rightarrow 0,\;N\gg 1
\end{equation}
while the leading term for the case of the complex Bethe roots reads 
\begin{equation}
\frak{p}(w_{N-1})\frak{p}(w_{N})\sim (zz^{\prime })^{\frac{4s}{r}%
-2}\prod_{n=1}^{N-2}w_{n}^{\frac{4s}{r}-2}=o(1)\;,\quad x\rightarrow
0,\;N\gg 1 .
\end{equation}
Thus we find for the two-string also that the arrangement of the real parts
of the Bethe roots is consistent with the Bethe equations in the ordered
limit. Note the absence of any holes. As indicated before, this is a
preliminary check on the string structure and a more detailed analysis will
be presented elsewhere \cite{KSP}.

\subsection{Spectral gaps of the leading excitations}

We now insert our results (\ref{Zh}) and (\ref{Zs}) on the zeroes of the
excitation spectrum into the general formula (\ref{Ex}). We recall that the
imaginary part of the zeroes is undetermined, but in the thermodynamic limit
becomes a continuous variable, say $\func{Im}u^{Z}=v\in \mathbb{R}$, which
parametrizes a band of eigenvalues corresponding to the excited state. In
the field theoretic context it can be identified with the rapidity. For the
hole-excitation we then find 
\begin{equation}
r_{\text{hole}}(u)=-\frac{\vartheta _{1}(\frac{\pi u}{6\lambda }-i\frac{\pi v%
}{6\lambda }-\frac{\pi }{6},|p|^{\frac{\pi }{6\lambda }})\vartheta _{1}(%
\frac{\pi u}{6\lambda }-i\frac{\pi v}{6\lambda }-\frac{\pi }{3},|p|^{\frac{%
\pi }{6\lambda }})}{\vartheta _{1}(\frac{\pi u}{6\lambda }-i\frac{\pi v}{%
6\lambda }+\frac{\pi }{6},|p|^{\frac{\pi }{6\lambda }})\vartheta _{1}(\frac{%
\pi u}{6\lambda }-i\frac{\pi v}{6\lambda }+\frac{\pi }{3},|p|^{\frac{\pi }{%
6\lambda }})}\;.  \label{rhole}
\end{equation}
In general the eigenvalues in the excitation band will be complex and in
order to determine the spectral gap, or equivalently the correlation length,
one has to integrate over the whole band. However, from the crossing
relation (\ref{cross}) we infer that the eigenvalues are real at the
isotropic point $u=3\lambda /2$. In addition, we have to tune the parameter $%
v$ such that we reach the bottom of the band. The latter value is derived to
be $v=\frac{\pi \tau }{12\lambda }$ and the correlation length is then
obtained via the formula 
\begin{equation}
\xi _{\text{hole}}(p)^{-1}=-\ln r_{\text{hole}}(\tfrac{3\lambda }{2})_{v=%
\frac{\pi \tau }{12\lambda }}=2\ln \frac{\vartheta _{4}(\frac{\pi }{4}+\frac{%
\pi }{3},|p|^{\frac{\pi }{6\lambda }})}{\vartheta _{4}(\frac{\pi }{4}-\frac{%
\pi }{3},|p|^{\frac{\pi }{6\lambda }})}\;.  \label{xih}
\end{equation}
Using a standard identity for the elliptic functions we obtain the series
expansion in the elliptic nome 
\begin{equation}
\xi _{\text{hole}}(p)^{-1}=4\sqrt{3}\sum_{n=0}^{\infty }\frac{(-1)^{n}}{6n+1}%
\,\frac{|p|^{\frac{\pi }{6\lambda }(6n+1)}}{1-|p|^{\frac{\pi }{3\lambda }%
(6n+1)}}+4\sqrt{3}\sum_{n=1}^{\infty }\frac{(-1)^{n}}{6n-1}\,\frac{|p|^{%
\frac{\pi }{6\lambda }(6n-1)}}{1-|p|^{\frac{\pi }{3\lambda }(6n-1)}},
\label{mh}
\end{equation}
from which we easily extract the amplitude (\ref{xi}) of the leading term in
the critical limit $p\rightarrow 0^{\pm }$, 
\begin{equation}
\mathcal{S}^{\pm }(0)^{-1}=\lim_{p\rightarrow 0^{\pm }}|p|^{-\frac{\pi }{%
6\lambda }}\xi _{\text{hole}}(p)^{-1}=4\sqrt{3}\;.  \label{xih0}
\end{equation}
Together with the general expression for the singular part of the free
energy (\ref{f0}) this puts us in a position to derive the universal
amplitude ratio (\ref{amp}): 
\begin{equation}
\mathcal{Q}^{\pm }(0)\mathcal{S}^{\pm }(0)^{2}=\frac{\sin \frac{\pi }{3}%
\frac{L}{L+2}}{4\sqrt{3}\sin \frac{2\pi }{3}\frac{L+1}{L+2}}\;,\quad L>3\;.
\label{ampT}
\end{equation}
This coincides with the result (\ref{tbaf}) from the thermodynamic Bethe
ansatz analysis reported in \cite{F} based on the exact scattering matrices
constructed in \cite{Sm} when the fundamental particle is related to a kink.
It needs to be pointed out that the general result (\ref{ampT}) agrees with
the outcome for the special cases $L=4,6$ previously reported in \cite{BS3}.
For $L=4$, i.e. for the leading thermal perturbation of the tricritical
Ising model, (\ref{ampT}) can also be compared against the numerical
findings obtained in \cite{FMS}.

The case $L=3$ is special as kinks are absent from the particle spectrum 
\cite{Sm}. In the context of the dilute $A_{3}$ model, also, numerical
investigations \cite{BNW,GN} of the Bethe equations did not show evidence
for the presence of a hole. As we have pointed out before, further
exceptions to the hole constituting the leading excitation might occur for $%
L $ even, when the sign of the elliptic nome is negated. This is, for
example, the case when $L=4$.

For these exceptional cases we therefore consider as the leading excitation
the two-string originally proposed in \cite{BNW} for $L=3$. In the context
of the scattering theory \cite{Sm} this type of excitation would correspond
to the lowest breather, which is the lightest particle exhibiting a
self-fusing process. From (\ref{Ex}) and (\ref{Zs}) we calculate the
associated spectral gap by repeating the same steps as before and find 
\begin{eqnarray}
\xi _{\text{string}}(p)^{-1} &=&2\ln \frac{\vartheta _{4}(\frac{\pi }{4}+%
\frac{\pi }{12}\frac{L-2}{L+2},|p|^{\frac{\pi }{6\lambda }})\vartheta _{4}(%
\frac{\pi }{4}+\frac{\pi }{12}\frac{3L+2}{L+2},|p|^{\frac{\pi }{6\lambda }})%
}{\vartheta _{4}(\frac{\pi }{4}-\frac{\pi }{12}\frac{L-2}{L+2},|p|^{\frac{%
\pi }{6\lambda }})\vartheta _{4}(\frac{\pi }{4}+\frac{\pi }{12}\frac{3L+2}{%
L+2},|p|^{\frac{\pi }{6\lambda }})}  \notag \\
&=&8\sum_{n=1}^{\infty }\frac{\sin \frac{n\pi }{2}}{n}\frac{|p|^{\frac{\pi }{%
6\lambda }n}}{1-|p|^{\frac{\pi }{3\lambda }n}}(\sin \tfrac{\pi n}{6}\tfrac{%
3L+2}{L+2}+\sin \tfrac{\pi n}{6}\tfrac{L-2}{L+2})  \notag \\
&\sim &8(\sin \tfrac{\pi }{6}\tfrac{3L+2}{L+2}+\sin \tfrac{\pi }{6}\tfrac{L-2%
}{L+2})|p|^{\frac{\pi }{6\lambda }},\;|p|\rightarrow 0.  \label{ms}
\end{eqnarray}
We point out that the sum (\ref{ms}) runs at most over the integers
relatively prime to 6, i.e. $n=1,5,7,11,...$ These integers coincide with
the allowed spins of integrals of motion compatible with a self-fusing
process \cite{FZ}. In order to obtain the universal amplitude (\ref{amp}) in
the case that the hole excitation is absent we need only to determine the
ratio of the two different correlation lengths in the critical limit, 
\begin{equation}
\frac{\xi _{\text{string}}(0)}{\xi _{\text{hole}}(0)}=\frac{\sin \frac{\pi }{%
3}}{\cos \tfrac{2\pi }{3(L+2)}+\sin \tfrac{\pi }{6}\tfrac{L-2}{L+2}}%
<1\;,\quad L>3.  \label{xir}
\end{equation}
As expected, we deduce that in presence of both a hole and a two-string the
smallest spectral gap is given by the hole-excitation. In the context of the
scattering matrices constructed in \cite{Sm} expression (\ref{xir})
determines the mass ratio of the kink and the lowest breather.

\section{The excitation spectrum for $L=3,4,6,\infty $}

For $L=3,4,6$ and regime 2 the excitation spectrum has been previously put
forward in the literature case-by-case, using the exact perturbation theory
approach in the ordered limit \cite{BS2, BS3, SB, SB3} and the different
string hypotheses based on numerical data \cite{BNW,GN,GNa,GNp,MO}. A
significant role in these proposals is played by the underlying Lie
algebraic structures related to the exceptional algebras $E_{8,7,6}$ which
also arise in the associated exact scattering theories.\smallskip

The excitation spectrum for the unrestricted model ($L=\infty $) connected
to the algebra $D_{4}$ has not been previously considered and its
formulation is a new result.\smallskip

We show that our proposal of a hole-excitation and a two-string excitation
leads to the correct Lie algebraic structures by connecting the location of
the real part of the zeroes (\ref{Zh}) and (\ref{Zs}) to the Coxeter
geometry of the respective algebras. Invoking similar Lie algebraic Coxeter
identities as were originally applied in the context of affine Toda field
theory \cite{PD,FO}, we summarize the previously reported excitation spectra
for $L=3,4,6$ in a single formula and prove additional functional equations
besides the ones already stated for general $L$. These relations completely
fix the location of the zeroes in (\ref{Ex}) for the higher fundamental
excitations \cite{BNW,MO,BS2,BS3,SB,SB3} as well, and can be regarded as
equivalent to a string hypothesis. We start by stating the excitation
spectrum in generic Lie algebraic terms for $L=3,4,6,\infty $ and then
identify the excitations belonging to the hole and the two-string.

\subsection{Excitations in terms of Coxeter geometry}

In the following let $\frak{g}$ stand for one of the simple Lie algebras $%
E_{8,7,6}$ or $D_{4}$. Then we expect, depending on the sign of the elliptic
nome, at most rank$\,\frak{g}$ fundamental excitations or eigenvalue bands
in the thermodynamic limit. Furthermore, we denote by $h$ the Coxeter number
of $\frak{g}$ which obeys the following identity in terms of the maximal
height value, 
\begin{equation}
h=6\;\frac{L+2}{L-2}\;.
\end{equation}
Introducing the simple roots $\alpha _{i}$ and the fundamental co-weights $%
\lambda _{i}^{\vee }$ of the Lie algebra $\frak{g}$ we define a matrix of
functions $\mu :\mathbb{Z}\rightarrow \frac{1}{2}\mathbb{Z}$ by setting 
\begin{equation}
\mu _{ij}(2a-\tfrac{c_{i}+c_{j}}{2})=-\frac{c_{i}}{2}\left\langle \lambda
_{j}^{\vee },\sigma ^{a}\alpha _{i}\right\rangle \;,\quad c_{i}=\pm 1\;.
\label{mu1}
\end{equation}
Here $\sigma $ denotes the Coxeter element of the Weyl group of $\frak{g}$.
It is uniquely defined w.r.t. to a specific bicolouration of the Dynkin
diagram specified by the ``colours'' $c_{i}$. See Appendix B for details. We
are now prepared to state the excitation spectrum for $L=3,4,6,\infty $: 
\begin{equation}
r_{j}(u)=(-1)^{A_{\star j}^{-1}}\prod\limits_{a=1}^{h}\left\{ \frac{%
\vartheta _{1}\left( \frac{h+2}{2h}u-\frac{\pi a}{2h}-i\frac{\pi \alpha }{2h}%
,|p|^{\frac{h+2}{2h}}\right) }{\vartheta _{1}\left( \frac{h+2}{2h}u+\frac{%
\pi a}{2h}-i\frac{\pi \alpha }{2h},|p|^{\frac{h+2}{2h}}\right) }\right\}
^{2\mu _{\star j}(a)}.  \label{Ex1}
\end{equation}
Here $j=1,2,...$rank$\,\frak{g}$ runs over the fundamental particle
excitations and the index $\star $ specifies the root in the Dynkin diagram
of the simple Lie algebra $\frak{g}$ which is connected to the affine root
in the Dynkin diagram of the affine extension $\frak{g}^{(1)}$ \cite{SB}. We
will show below that this particular node is singled out by the two-string.
The variable $\alpha \in \mathbb{R}$ parametrizes the imaginary part of the
Bethe roots in the thermodynamic limit and scales the eigenvalue bands. The
sign factor in front is determined by the inverse Cartan matrix which is
given in terms of the simple roots as $A_{kl}=2\left\langle \alpha
_{k},\alpha _{l}\right\rangle /\left\langle \alpha _{k},\alpha
_{k}\right\rangle $. In fact, the Cartan matrix allows for an alternative
definition of the exponent function (\ref{mu1}), which will be of great
practical use in the subsequent computations. Consider the $q$-deformed
Cartan matrix $[A]_{q}$ defined by 
\begin{equation}
\left( \lbrack A]_{q}\right) _{kl}:=[A_{kl}]_{q}\;,\quad \lbrack n]_{q}:=%
\frac{q^{n}-q^{-n}}{q-q^{-1}}.
\end{equation}
Then the functions (\ref{mu1}) can be implicitly defined by the relation 
\cite{FKS,thesis} 
\begin{equation}
\frac{1-q^{2h}}{2}[A]_{q}^{-1}=\sum_{a=1}^{2h}\mu (a)q^{a}\;.  \label{mu2}
\end{equation}
The above relation is non-trivial as it states that the inverse $q$-deformed
Cartan matrix, which generically is a rational function in $q$, simplifies
to a polynomial when multiplied with the factor $(1-q^{2h})$. The functions $%
\mu $ then simply count how often the monomial $q^{a}$ occurs in this
expansion. That the two definitions of $\mu $ are indeed equivalent has been
proven in \cite{FKS,thesis}.\emph{\ }One can also now check explicitly that
the above expression (\ref{Ex1}) correctly reproduces the case-by-case
results derived previously \cite{BS2, BS3, SB, SB3}\footnote{%
Setting $\alpha =\frac{h+2}{2\pi }\tau $ we obtain the leading contribution
previously reported in the literature.}. There and in \cite{MO} it was
noticed that the characteristic integers $a$ fixing the real part of the
zeroes in the excitation spectrum (\ref{Ex1}) match the integers appearing
in the building blocks of hyperbolic functions used in describing the $ADE$
affine Toda S-matrices \cite{BCDS}. In the context of affine Toda field
theory the relation of these integers to Coxeter geometry was established in 
\cite{PD,FO}.

One can now explicitly check that the resulting zeroes (\ref{Zs}) and (\ref
{Zh}) from the hole and two-string excitation match the Lie algebraic
formula (\ref{Ex1}) when the node $\star $ is specified by the simple root
connected to the affine root of $\frak{g}$. Employing the helpful identities 
\begin{equation*}
\lambda =\frac{\pi }{2r}\,\frac{h}{3}\quad \text{and\quad }2r=2\Delta
_{1,2}^{-1}=h+2
\end{equation*}
one finds the real parts of the zeroes listed in Table 3 (a) and 3 (b). We
note that the hole gives always the lightest particle in the mass spectrum
(except for $L=3$) while the two-string associated with the node $\star $
corresponds to the lightest particle which exhibits a self-fusing process (%
\ref{Sboot}). It is remarkable that from the simple excitation of a hole and
a two-string one already sees the four different Lie algebraic structures
emerging.\medskip

\begin{center}
\begin{tabular}{|l|l|l|}
\hline\hline
$L$ & algebra & $\frac{h+2}{\pi }\,\func{Re}u^{Z}$ \\ \hline\hline
$4$ & $E_{7}$ & $6,12$ \\ \hline\hline
$6$ & $E_{6}$ & $4,8$ \\ \hline\hline
$\infty $ & $D_{4}$ & $2,4$ \\ \hline\hline
\end{tabular}
\quad \quad \quad 
\begin{tabular}{|l|l|l|l|}
\hline\hline
$L$ & algebra & $\star $ & $\frac{h+2}{\pi }\,\func{Re}u^{Z}$ \\ \hline\hline
$3$ & $E_{8}$ & 1 & $1,11,19,29$ \\ \hline\hline
$4$ & $E_{7}$ & 2 & $1,7,11,17$ \\ \hline\hline
$6$ & $E_{6}$ & 2 & $1,5,7,11$ \\ \hline\hline
$\infty $ & $D_{4}$ & 2 & $1,3^{2},5$ \\ \hline\hline
\end{tabular}
\smallskip

{\small (a)\quad \quad \quad \quad \quad \quad \quad \quad \quad \quad \quad
\quad \quad \quad \quad \quad \quad (b)}\medskip
\end{center}

\noindent {\small \textbf{Table 3 (a) and 3 (b)}. Real parts of the zeroes
in units }$\frac{\pi }{h+2}${\small \ describing the excitation spectrum
corresponding to the presence of a hole (a) and the two-string (b). The
results are in accordance with the general formula (\ref{Ex1}) and display
the Coxeter geometry of the underlying Lie algebra. Note that the hole
excitation corresponding to the kink is absent for L=3. For L\TEXTsymbol{>}3
this excitation corresponds to the particle index j=1. The numeration of the
nodes is in accordance with the conventions in \cite{BS2, BS3, SB, SB3}.}%
\medskip

The excitations (\ref{Ex1}) with particle indices $j>2$ ought to belong to
heavier particles which can be expressed as bound states of the former ones.
This suggest that the corresponding Bethe roots distributions, which might
contain additional strings or holes, have to fulfill further constraints
originating from bootstrap identities other than (\ref{rboot}). At the
moment we leave further exploration of the string structure of the dilute $%
A_{L}$ models to future work \cite{KSP} and simply state these conjectured
excitations as additional consistent solutions to the equation (\ref{rboot})
which naturally arise from the Lie algebraic structure. Nevertheless, as a
preparatory step for a closer investigation of the string structure, we
state the additional bootstrap identities arising from Coxeter geometry in
the next section.

\subsubsection{Bootstrap identities}

We start by verifying that the explicit expression (\ref{Ex1}) for $%
L=3,4,6,\infty $ obeys the functional relations (\ref{HA}), (\ref{cross})
and (\ref{rboot}). To this end it is helpful to introduce the meromorphic
function 
\begin{equation}
t(u,a):=\frac{\vartheta _{1}\left( \frac{h+2}{2h}u-\frac{\pi a}{2h},|p|^{%
\frac{h+2}{2h}}\right) }{\vartheta _{1}\left( \frac{h+2}{2h}u+\frac{\pi a}{2h%
},|p|^{\frac{h+2}{2h}}\right) }  \label{t}
\end{equation}
which enjoys the following transformation properties 
\begin{eqnarray}
t(u+6\lambda ,a) &=&t(u,a\pm 2h)=t(-u,a)^{-1}=t(u,-a)^{-1}=t(u,a)  \label{ta}
\\
t(u+3\lambda ,a) &=&-t(u,h-a)^{-1}  \label{tb} \\
t(u+i\tau ,a) &=&e^{2\pi i\frac{a}{h}}t(u,a)\;.  \label{tc}
\end{eqnarray}
Using the above identities the verification of the functional equations (\ref
{HA}), (\ref{cross}) and (\ref{rboot}) becomes a purely algebraic problem
which can be formulated entirely in terms of the generating function (\ref
{mu1}). The latter has been thoroughly investigated in the context of affine
Toda field theory \cite{FO,FKS,thesis} and proceeding in an analogous manner
one can establish a one-to-one match between the functional relations
satisfied by the excitation spectrum of the dilute $A_{3,4,6,\infty }$ model
and the functional identities obeyed by the factorizable scattering matrices
of the corresponding affine Toda field theory.

The properties of the generating function (\ref{mu1}) which we exploit are 
\cite{FO,FKS,thesis} 
\begin{equation}
\mu (a)=\mu (2h+a)=-\mu (2h-a)\;.  \label{id1}
\end{equation}
An additional remarkable identity which is not straightforward to prove
reads \cite{FO,FKS,thesis} 
\begin{equation}
\mu _{kl}(h-a)=\mu _{\bar{k}l}(a),  \label{idc}
\end{equation}
where $\bar{k}$ denotes the node in the Dynkin diagram of the Lie algebra $%
\frak{g}$ which is obtained by applying a possible Dynkin diagram
automorphism to the node $k$. The second definition of the generating
function (\ref{mu2}) is most useful for verifying the formulas 
\begin{equation}
\mu (a)=\mu (a)^{t}\text{\quad and\quad }\mu (a+1)+\mu (a-1)=(2-A)\mu (a)\,.
\label{id2}
\end{equation}

We note that the relation (\ref{HA}) corresponding to hermitian analyticity
in exact S-matrix theory is trivially satisfied due to (\ref{ta}). The
inversion (\ref{rinv}) and crossing relation (\ref{cross}) then follow by
observing that the preferred node $\star $ is self-conjugate, i.e. $\star =%
\bar{\star}$, and employing (\ref{tb}). The number of zeroes is given by the
inverse Cartan matrix element $A_{k\star }^{-1}$, which in the case of the $%
E $-type algebras coincides with the Kac or Dynkin labels: 
\begin{equation}
A_{k\star }^{-1}=\sum_{a=1}^{h}2\mu _{\star k}(h-a)\frac{a}{h}%
=\sum_{a=1}^{h}2\mu _{\star k}(a)\frac{a}{h}\;.
\end{equation}
From this algebraic identity together with (\ref{tc}) we now also read off
the double-periodicity (\ref{period}) of the excitation spectrum (\ref{Ex1}).

The central equation (\ref{rboot}) requires a deeper algebraic analysis
using Coxeter geometry. In fact, it is a special case of more general
identities, which, along the lines of the analogous analysis in affine Toda
field theory \cite{PD,FO}, can be formulated as follows (see Appendix B on
Coxeter geometry for the explanation of the various symbols):

Whenever there exist three elements in the Coxeter orbits $\Omega _{l},$
associated with the simple roots $\gamma _{l}=c_{l}\alpha _{l},\;c_{l}=\pm 1$
and $l=i,j,k$ which sum to zero, i.e. 
\begin{equation*}
\sum_{l=i,j,k}\sigma ^{\xi _{l}}\gamma _{l}=0\quad \Leftrightarrow \quad
\sum_{l=i,j,k}\mu (a\pm \eta _{l})=0\;,\quad \eta _{l}:=-2\xi _{l}
\end{equation*}
for some triplet of integers $\xi _{l}$, one has the identity 
\begin{equation}
\prod_{l=i,j,k}r_{l}(u+3\lambda \tfrac{\eta _{l}}{h})=1.  \label{boot}
\end{equation}
For $i=j=k$ we recover equation (\ref{rboot}) setting $\eta _{l}=0,2h/3,4h/3$%
. In the context of affine Toda field theory these ``bootstrap'' identities
occur when an intermediate bound state is formed, $i+j\rightarrow \bar{k}$
with $\bar{k}$ denoting the antiparticle of $k$. We see that in general
these identities involve different particle indices. A few examples are
listed in Table 4.\medskip

\begin{center}
\begin{tabular}{|l|l|l|}
\hline\hline
$L$ & Coxeter identity & bootstrap equation \\ \hline\hline
$3$ & $\alpha _{2}+\sigma ^{13}\alpha _{3}+\sigma ^{23}\alpha _{1}=0$ & $%
r_{2}(u+\frac{23}{5}\lambda )r_{3}(u+2\lambda )r_{1}(u)=1$ \\ \hline\hline
$4$ & $\alpha _{5}+\sigma ^{7}\alpha _{7}+\sigma ^{14}\alpha _{4}=0$ & $%
r_{5}(u+\frac{14}{3}\lambda )r_{7}(u+\frac{7}{3}\lambda )r_{4}(u)=1$ \\ 
\hline\hline
$6$ & $\alpha _{1}+\sigma ^{5}\alpha _{3}+\sigma ^{10}\alpha _{2}=0$ & $%
r_{1}(u+5\lambda )r_{3}(u+\frac{5}{2}\lambda )r_{2}(u)=1$ \\ \hline\hline
\end{tabular}
\medskip
\end{center}

\noindent {\small \textbf{Table 4}. Displayed are three bootstrap identities
for the excitations of the dilute }$A_{3,4,6}$ {\small models and the
corresponding Coxeter identities in the root systems of the associated }$%
E_{8,7,6}$ {\small algebras. The numeration of the nodes is in accordance
with the convention in \cite{BS2, BS3, SB, SB3}.}\medskip\ 

In the present context, these bootstrap identities completely fix the
position of the poles and zeroes and are therefore equivalent to a string
hypothesis. Only the identity (\ref{rboot}) is common to all models; the
remaining bootstrap equations characterize the dilute $A_{L}$ model for
different $L=3,4,6,\infty $. They have a purely geometrical origin in the
structure of the underlying algebra's root system and we conclude that just
as the bound state structure of affine Toda field theory can be encoded in
Coxeter geometry, so can the string structure of the dilute $A_{3,4,6,\infty
}$ models. The underlying Lie algebra also becomes manifest in another
identity \cite{FKS,thesis} which can be built up using the more
``fundamental'' bootstrap identities (\ref{boot}): 
\begin{equation}
r_{j}(u+\tfrac{\pi }{h+2})r_{j}(u-\tfrac{\pi }{h+2})=\prod_{l=1}^{\text{rank}%
\frak{g}}r_{l}(u)^{(2-A)_{jl}}\;.  \label{cboot}
\end{equation}
Here $A$ is the Cartan matrix of the particular Lie algebra in question.
This identity directly reflects the Dynkin diagram structure. Like the more
general bootstrap identities, (\ref{boot}) it points towards a relation
between strings belonging to different excitations which we will investigate
in a forthcoming publication \cite{KSP}.

\subsubsection{Spectral gaps and affine Toda mass spectra}

The excitation spectrum (\ref{Ex1}) explicitly states that in regime 2 there
are at most 8,7,6 or 4 continuous bands of eigenvalues of the transfer
matrix for $L=3,4,6,\infty $, respectively. Each band belonging to one of
the leading excitations is separated from the groundstate by a gap. We now
calculate the associated critical amplitudes and corrections to the scaling
behavior (\ref{xi}) for all conjectured fundamental excitations (\ref{Ex1}).
It will be worthwhile to do this in the framework of Coxeter geometry in
order to reveal the Lie algebraic structure in the scaling corrections and
relate them to the affine Toda mass spectra.

As before we evaluate the excitations at the isotropic point $u=3\lambda /2$
where the eigenvalues are real and set the parameter $\alpha $ in (\ref{Ex1}%
) to the value $\alpha _{0}=(h+2)\tau /\pi $ giving the lowest eigenvalue in
each band. We then find 
\begin{equation}
\xi _{k}(p)^{-1}=-\ln r_{k}(\tfrac{3\lambda }{2})_{\alpha =\frac{h+2}{\pi }%
\tau }=\sum_{a=1}^{h}2\mu _{\star k}(a)\ln \frac{\vartheta _{4}\left( \frac{%
\pi }{4}+\frac{\pi a}{2h},|p|^{\frac{h+2}{2h}}\right) }{\vartheta _{4}\left( 
\frac{\pi }{4}-\frac{\pi a}{2h},|p|^{\frac{h+2}{2h}}\right) }\;.
\end{equation}
Since we are interested in the behaviour of the correlation length in the
critical limit, it is favorable to rewrite the above expression in terms of
a power series in the elliptic nome using a standard identity for the
logarithm of Jacobi's theta functions, 
\begin{equation}
\xi _{k}(p)^{-1}=4\sum_{n=1}^{\infty }\frac{\sin \frac{\pi n}{2}}{n}\frac{%
|p|^{\frac{h+2}{2h}n}}{1-|p|^{\frac{h+2}{h}n}}\sum_{a=1}^{h}2\mu _{\star
k}(a)\sin \frac{\pi a}{h}n\;.  \label{m1a}
\end{equation}
From the coefficients in this expansion, which contain the Coxeter integers,
we need now to extract the affine Toda masses. This has been done in the
literature \cite{BS3} for $L=3,4,6$ case-by-case using trigonometric
identities. It is worthwhile to review these calculations using Coxeter
geometry to reveal the underlying Lie algebraic structure. To this end we
note the following formula for the residue of the inverse $q$-deformed
Cartan matrix: 
\begin{equation}
\left. \limfunc{Res}\left[ A\right] _{q=e^{x}}^{-1}\right| _{\frac{i\pi s}{h}%
}=\frac{(-1)^{s}i}{h}\sum_{a=1}^{h}2\mu (a)\sin \frac{\pi a}{h}s=-\frac{%
iP^{s}}{2\sin \frac{\pi s}{h}}\;.  \label{ResA}
\end{equation}
Here $1\leq s\leq h-1$ is an exponent of the Lie algebra in question and we
have used the identity (\ref{mu2}). The second equality in (\ref{ResA})
involves the matrix 
\begin{equation}
P^{s}=\frac{1}{2\pi i}\oint\limits_{|\zeta -2\cos \frac{\pi s}{h}%
|=\varepsilon }d\zeta \;(\zeta -2+A)^{-1}
\end{equation}
which is the orthogonal projector onto the eigenspaces of the (non-deformed)
Cartan matrix, 
\begin{equation}
(P^{s})^{2}=P^{s}\quad \text{and\quad }AP^{s}=4\sin ^{2}\frac{\pi s}{2h}%
\,P^{s}\;.
\end{equation}
The matrix elements of the orthogonal projector consist of the components of
the normalized eigenvectors. Setting $s=1$ we obtain the\ components of the
Perron-Frobenius eigenvector which yields the (classical) affine Toda masses 
\cite{FLO} 
\begin{equation}
P_{kl}^{s=1}=m_{k}m_{l}\;.
\end{equation}
Since all eigenvalues of the Cartan matrix are given in terms of the
exponents, we conclude that the residue (\ref{ResA}) vanishes if $s$ is an
integer which does not belong to the set of exponents. The final result for
the correlation length and the leading term in the critical limit $%
p\rightarrow 0$ therefore reads 
\begin{equation}
\xi _{k}(p)^{-1}=\sum_{\hat{s}}\frac{2h}{\hat{s}}\frac{\sin \frac{\pi \hat{s}%
}{2}}{\sin \frac{\pi \hat{s}}{h}}\,P_{\star k}^{\hat{s}}\frac{|p|^{\frac{h+2%
}{2h}\hat{s}}}{1-|p|^{\frac{h+2}{h}\hat{s}}}\sim 2h\,\frac{m_{\star }m_{k}}{%
\sin \frac{\pi }{h}}\,|p|^{\frac{h+2}{2h}}+...  \label{m1}
\end{equation}
Here $\hat{s}$ runs over the affine exponents, i.e. the finite exponents $s$
modulo multiples of the Coxeter number, $\hat{s}=s\func{mod}h$. They are
given in Table 5.\medskip

\begin{center}
\begin{tabular}{|l|l|}
\hline\hline
Algebra & exponents $s$ \\ \hline\hline
$E_{8}$ & $1,7,11,13,17,19,23,29$ \\ \hline\hline
$E_{7}$ & $1,5,7,9,11,13,17$ \\ \hline\hline
$E_{6}$ & $1,4,5,7,8,11$ \\ \hline\hline
$D_{4}$ & $1,5,7,11$ \\ \hline\hline
\end{tabular}
\medskip
\end{center}

\noindent {\small \textbf{Table 5}. The exponents of particular finite Lie
algebras, which determine the eigenvalues of the corresponding Cartan matrix.%
}\medskip

\noindent Note that only the odd exponents contribute to the sum in (\ref{m1}%
) and that we have chosen a normalization for the lightest mass different
from $m_{1}=1$. Together with the result for the free energy density (\ref
{f0}) the universal quantity (\ref{amp}) is expressed as 
\begin{equation}
\mathcal{Q}^{\pm }(0)\mathcal{S}^{\pm }(0)^{2}=4\sqrt{3}\frac{\cos \pi \frac{%
h-3}{3h}}{\cos \frac{\pi }{h}}\left( \frac{\sin \frac{\pi }{h}}{2h\,m_{\star
}m_{1}}\right) ^{2}\;.  \label{todaAmp}
\end{equation}
The higher powers of the elliptic nome in the scaling functions (\ref{m1})
originate from irrelevant operators of the minimal CFT $M_{L,L+1}$ and the
specific choice of the scaling variables. For the Ising model in a magnetic
field, both contributions to the renormalization behavior have been analyzed
in \cite{CH}. We therefore briefly comment on the scaling corrections in
this case.

\paragraph{The scaling functions for $L=3$.}

Setting $L=3$ in (\ref{m1}) we explicitly write out the first scaling
corrections to the spectral gap of the dilute $A_{3}$ model \cite{SB0}, 
\begin{equation}
|p|^{-\frac{8}{15}}\xi _{k}^{-1}=\mathcal{S}^{\pm }(0)^{-1}\left( 1+|p|^{%
\frac{16}{15}}+|p|^{\frac{32}{15}}+S_{1}|p|^{\frac{48}{15}}+...\right) \;.
\end{equation}
In comparison with \cite{CH} we note the absence of the analytic terms. This
absence might be explained by a different choice of scaling variables, i.e.
the identification of the elliptic nome with the magnetic field in the Ising
model differs from \cite{CH}. One has to keep in mind that the dilute $A_{L}$
model is not necessarily equivalent to the Ising model at the critical point
but only resides in the same universality class. In contrast, the terms
depending on the irrelevant operators of the identity family in $M_{3,4}$
are present \cite{CH}. The latter are related to the components $T\bar{T}%
,T^{2},\bar{T}^{2}$ of the energy momentum tensor.

From the expansion (\ref{f}) we may also compare the scaling corrections for
the singular part of the free energy density with those of \cite{CH}, 
\begin{equation*}
|p|^{-\frac{16}{15}}f_{s}=\mathcal{Q}^{\pm }(0)\left( 1+Q_{1}|p|^{\frac{16}{%
15}}+Q_{2}|p|^{\frac{32}{15}}+Q_{3}|p|^{\frac{48}{15}}+...\right)
\end{equation*}
Once more we find all terms originating from the irrelevant operators
belonging to the class of the identity field reported in \cite{CH}.

\section{Conclusions}

In this article we have put forward the leading excitations of the dilute $%
A_{L}$ model in regime 2 as consisting of a hole and a two-string. Our
proposal is for special values of $L$ supported by numerical computations
and further support comes from revealing the expected Lie algebraic
structures $E_{8,7,6}$ and $D_{4}$ for $L=3,4,6$ and $L=\infty $, i.e. the
unrestricted model. A crucial role in exhibiting these structures is played
by analyticity arguments and Coxeter geometry which have allowed us to
relate the zeroes in the excitation spectrum induced by the Bethe roots
directly to the underlying Lie algebra and to cast the excitation spectrum
of all four cases into a single formula (\ref{Ex1}). This procedure is more
direct than using the non-linear integral equations of the thermodynamic
Bethe ansatz, where one relates the corresponding integral kernels to the $q$%
-deformed Cartan matrix in Fourier space as has been done for $L=3$ \cite
{BNW}. We note that the excitation spectrum (\ref{Ex1}) for $E_{8}$ in fact
matches the thermodynamic Bethe ansatz outcome of \cite{BNW} and the result
of the exact perturbation theory approach \cite{BS2} but we leave a closer
investigation of the proposed string structure for the higher excitations to
future work \cite{KSP}. We believe that the one-to-one correspondence
between the functional equations (\ref{boot}) and the bootstrap identities
of the respective scattering matrices will be of central significance in
gaining a thorough understanding of the allowed strings and a possible
classification of the excitations.

Our result for the spectral gaps and the critical amplitudes connected with
the correlation length can now be compared against the result (\ref{tbaf})
from the thermodynamic Bethe ansatz analysis of the relativistic scattering
matrices. We briefly report the result for the cases $L=3,4,6$ where the
associated scattering matrices can be written in the following universal
formula (see e.g. \cite{FKS,thesis} for a derivation): 
\begin{equation}
S(\beta )=\exp \int_{0}^{\infty }\frac{dt}{t}\,4\cosh t\,\left( \left[ A%
\right] _{q=e^{t}}^{-1}\right) _{11}\sinh \frac{h\beta }{i\pi }t\;.
\end{equation}
We note that the $q$-deformed Cartan matrix $\left[ A\right] _{q}$ which we
frequently used in our analysis of the transfer matrix excitation spectrum
also appears here. A similar calculation to that we have performed in
section 4 in the context of elliptic functions, yields upon invoking formula
(\ref{tbaf}) 
\begin{equation}
\mathcal{Q}^{\pm }(0)\mathcal{S}^{\pm }(0)^{2}=\frac{\tan \frac{\pi }{h}}{%
4h\,m_{1}^{2}}\;,
\end{equation}
which is found to coincide with our result (\ref{todaAmp}). Here $m_{1}$
denotes the smallest component of the Perron-Frobenius eigenvector of the
Cartan matrix, which can be identified with the classical mass. For general
height values $L$ our result (\ref{ampT}) confirms the result in \cite{F}
based on the thermodynamic Bethe ansatz analysis of Smirnov's scattering
matrices \cite{Sm} when the hole excitation is identified with the presence
of a kink state. As the universal amplitude enters directly into the bulk
vacuum expectation value of the energy-momentum tensor needed in the form
factor approach to construct correlation functions, this underlines the
intimate relationship between integrable lattice models and field theories.

Another field theoretic aspect which needs to be mentioned is the
interpretation of the particle spectra in light of the conformal structure
at the critical point. For the special cases $L=3,4,6$ where an underlying
exceptional Lie algebra is present, the particle spectra manifest themselves
in Rogers-Ramanujan identities for the conformal characters \cite
{KKMM,WP,WPb,MO}. Whether similar identities can be found for general $L$
and further information on the conformal field theory can be extracted from
the excitation spectrum are open challenges.

We close by pointing out that further universal amplitude ratios can be
derived from the critical amplitudes we have computed in this article. For
example, provided that $L$ is even the elliptic nome can be identified with
the reduced temperature up to a metric factor, $p\equiv -t=1-T/T_{c}$. This
allows one to compute the following universal quantity involving the
(singular) specific heat $C_{s}$ \cite{PHA}, 
\begin{equation}
R_{\xi }^{\pm }=\mathcal{A}^{\frac{1}{2}}\mathcal{S}^{\pm }(0)\;,\quad
C_{s}=-\frac{\partial ^{2}f_{s}}{\partial t^{2}}\sim \frac{\mathcal{A}}{%
\alpha }\,t^{-\alpha }\;.  \label{R}
\end{equation}
Employing the concept of hyperscaling, the critical exponent $\alpha $
controlling the divergence of the specific heat near the critical point can
be related to the critical exponent controlling the divergence of the
correlation length: 
\begin{equation}
2\nu =2-\alpha =\left( 1-\Delta _{1,2}\right) ^{-1}=\frac{4}{3}\frac{L+1}{L+2%
}\;,\quad L\text{ even\ .}
\end{equation}
The coefficient $\mathcal{A}$ in (\ref{R}) is easily obtained from the
expansion of the singular free energy density (\ref{f0}) to be 
\begin{equation}
\mathcal{A}=4\sqrt{3}\,\alpha (1-\alpha )(2-\alpha )\frac{\sin \frac{\pi }{3}%
\frac{L}{L+2}}{\sin \frac{2\pi }{3}\frac{L+1}{L+2}}\;.
\end{equation}
The important observation in this context is that for $L$ even the critical
amplitudes $\mathcal{S}^{\pm }(0)$ of the correlation length above ($p<0$)
and below ($p>0$) the critical temperature do not necessarily coincide. This
is indeed the case for $L=4$, i.e. the leading thermal perturbation of the
tricritical Ising model, where the particle corresponding to the hole
excitation is absent for $p>0$ and the heavier particle associated with the
two-string gives the leading contribution \cite{SB3}. The different
universal amplitudes above and below the critical temperature are related
through the ratio (\ref{xir}) which for $L=4$ coincides with the previously
reported result \cite{BS3,SB3,FMS} 
\begin{equation}
\mathcal{S}^{+}(0)/\mathcal{S}^{-}(0)=1/2\cos \tfrac{5\pi }{18}\;.
\end{equation}
The absence of the hole excitation in depending on the sign of the elliptic
nome should follow from symmetry arguments similar to those in the field
theoretic context \cite{lmc,M}. To see this directly from the Bethe ansatz
is an interesting problem for future investigations and requires a deeper
analysis of the Bethe equations of the dilute $A_{L}$ model in the
thermodynamic limit and of the string structure of the excited states.
Finally, we expect that the analysis of the excitation spectrum via
functional equations as outlined in this work can also be applied to the
other regimes of the dilute $A_{L}$ model, though in a possibly modified
form. The relevant universal amplitude ratios (\ref{amp}) and (\ref{R}) for
regime 1, corresponding to the $\phi _{2,1}$ perturbation, have recently
been reported in \cite{KAS}.{\small \medskip }

\noindent \textbf{Acknowledgments}. It is a pleasure to thank Andreas Fring,
Olaf Lechtenfeld, Barry McCoy, Will Orrick and Itzhak Roditi for interesting
discussions and comments. KAS also wishes to acknowledge related
correspondence with Patrick Dorey.\ This collaboration started during the
sabbatical leave of KAS at the C.N. Yang Institute and we are especially
thankful to Barry McCoy for his support and continuing interest in this
work. CK is financially supported by the Research Foundation Stony Brook,
NSF Grants DMR-0073058 and PHY-9988566.

\appendix

\section{The eigenspectrum after the conjugate modulus transformation}

We briefly sketch how to prove the asymptotic behavior (\ref{asym}) in the
ordered limit $|p|\rightarrow 1$. Exploiting the conjugate modulus
representation (\ref{E}) of Jacobi's theta functions the eigenvalue spectrum
(\ref{eig}) can be rewritten as 
\begin{multline*}
(-e^{2u^{2}/\tau }w^{\frac{3s}{r}})^{N}\Lambda (w)=\omega \left[ \frac{%
E(x^{4s}/w)E(x^{6s}/w)}{E(x^{4s})E(x^{6s})}\right] ^{N}%
\prod_{j=1}^{N}w_{j}^{1-\frac{2s}{r}}\frac{E(x^{2s}w/w_{j})}{E(x^{2s}w_{j}/w)%
} \\
+\left[ \frac{x^{2s}E(w)E(x^{6s}/w)}{wE(x^{4s})E(x^{6s})}\right]
^{N}\prod_{j=1}^{N}\frac{w_{j}E(w/w_{j})E(x^{6s}w_{j}/w)}{%
E(x^{2s}w_{j}/w)E(x^{4s}w_{j}/w)} \\
+\omega ^{-1}\left[ \frac{^{{}}wE(w)E(x^{2s}/w)}{x^{6s-2r}E(x^{4s})E(x^{6s})}%
\right] ^{N}\prod_{j=1}^{N}w_{j}^{\frac{2s}{r}-1}\frac{E(x^{8s-2r}w_{j}/w)}{%
E(x^{4s}w_{j}/w)}\;.
\end{multline*}
Here we have introduced the variables $w=e^{-\frac{2\pi }{\tau }u},\;s=\frac{%
L+2}{L-2}$ and $x=e^{-\frac{\pi ^{2}}{\tau r}}$ with $r=4\frac{L+1}{L-2}$ in
regime 2 with $p>0$. The nome $q^{2}=x^{2r}$ is suppressed in the notation.
(Note that the parameters $r$ and $s$ and hence the variable $x$ have been
rescaled from those used by the authors of \cite{BS} in their various
papers.) Assuming that for the groundstate all Bethe roots lie on the unit
circle we derive for $0<\tau \ll 1$ ($x\ll 1$), 
\begin{equation*}
\frac{e^{\frac{2u^{2}N}{\tau }}\Lambda _{0}(w)}{(-w^{-\frac{3s}{r}})^{N}}%
\sim \left\{ 
\begin{array}{cc}
\omega _{0}\prod\limits_{n}w_{n}^{1-\frac{2s}{r}}+(\frac{x^{2s}}{w}%
)^{N}\prod\limits_{n}w_{n}+\omega _{0}^{-1}x^{2sN}\prod\limits_{n}w_{n}^{%
\frac{2s}{r}}, & \frac{\pi }{L+1}<u<\lambda \\ 
\omega _{0}(\frac{w}{x^{2s}})^{N}\prod\limits_{n}w_{n}^{-\frac{2s}{r}%
}+1+\omega _{0}^{-1}(\frac{x^{4s}}{w})^{N}\prod\limits_{n}w_{n}^{\frac{2s}{r}%
}, & \lambda <u<2\lambda \\ 
\omega _{0}x^{2sN}\prod\limits_{n}w_{n}^{-\frac{2s}{r}}+(\frac{w}{x^{4s}}%
)^{N}\prod\limits_{n}w_{n}+\omega _{0}^{-1}\prod\limits_{n}w_{n}^{\frac{2s}{r%
}-1}, & 2\lambda <u<3\lambda
\end{array}
\right.
\end{equation*}
and 
\begin{equation*}
\frac{e^{\frac{2u^{2}N}{\tau }}\Lambda _{0}(w)}{(-w^{-\frac{3s}{r}})^{N}}%
\sim \omega _{0}\prod\limits_{n}w_{n}^{1-\frac{2s}{r}}+(\tfrac{x^{2s}}{w}%
)^{N}\prod\limits_{n}w_{n}+\omega _{0}^{-1}(\tfrac{w}{x^{6s-2r}}%
)^{N}\prod\limits_{n}w_{n}^{\frac{2s}{r}-1},\quad 0<u<\tfrac{\pi }{L+1}\;.
\end{equation*}
We therefore have the following asymptotic behaviour as $N\rightarrow \infty 
$: 
\begin{equation*}
\frac{e^{\frac{2u^{2}N}{\tau }}\Lambda _{0}(w)}{(-w^{-\frac{3s}{r}})^{N}}%
\sim \left\{ 
\begin{array}{cc}
\omega _{0}\prod\limits_{n}w_{n}^{1-\frac{2s}{r}}, & 0<\func{Re}u<\lambda \\ 
1, & \lambda <\func{Re}u<2\lambda \\ 
\omega _{0}^{-1}\prod\limits_{n}w_{n}^{\frac{2s}{r}-1}, & 2\lambda <\func{Re}%
u<3\lambda
\end{array}
\right. \;.
\end{equation*}
We thus conclude that in the thermodynamic limit $N\gg 1$, 
\begin{equation*}
\Lambda _{0}(u)\sim \left\{ 
\begin{array}{cc}
\omega _{0}\left\{ \frac{\vartheta _{1}(2\lambda -u)\vartheta _{1}(3\lambda
-u)}{\vartheta _{1}(2\lambda )\vartheta _{1}(3\lambda )}\right\} ^{N}G_{0}(u)%
\left[ 1+o(e^{-N})\right] , & 0<u<\lambda \\ 
\left\{ \frac{\vartheta _{1}(u)\vartheta _{1}(3\lambda -u)}{\vartheta
_{1}(2\lambda )\vartheta _{1}(3\lambda )}\right\} ^{N}\frac{G_{0}(u-\lambda )%
}{G_{0}(u-2\lambda )}\left[ 1+o(e^{-N})\right] , & \lambda <u<2\lambda \\ 
\omega _{0}^{-1}\left\{ \frac{\vartheta _{1}(u)\vartheta _{1}(\lambda -u)}{%
\vartheta _{1}(2\lambda )\vartheta _{1}(3\lambda )}\right\}
^{N}G_{0}(u-3\lambda )^{-1}\left[ 1+o(e^{-N})\right] , & 2\lambda <u<3\lambda
\end{array}
\right. \;.
\end{equation*}
Since the excited states $\Lambda (u)$ differ only by a finite number of
Bethe roots with non-vanishing real part we obtain analogous asymptotic
behaviour in the thermodynamic limit for them also. This proves (\ref{asym})
and therefore the functional equation (\ref{Boot0}).

\section{Coxeter geometry}

Let $\frak{g}$ denote in the following a simple Lie algebra and denote by $%
\{\alpha _{1},...,\alpha _{n}\}$ a set of simple roots which span its root
system, i.e. the eigenvalues of the Cartan subalgebra in the adjoint
representation. Each simple root $\alpha _{i}$ can be interpreted as element
in an Euclidean vector space $\mathbb{R}^{n}$ of dimension $n=$rank$\,\frak{g%
}$. It naturally defines a reflection $\mathbb{R}^{n}\rightarrow \mathbb{R}%
^{n}$ at its associated hyperplane by setting 
\begin{equation}
v\rightarrow \sigma _{i}v:=v-\frac{2\left\langle v,\alpha _{i}\right\rangle 
}{\left\langle \alpha _{i},\alpha _{i}\right\rangle }\,\alpha _{i}\;,\quad
v\in \mathbb{R}^{n}.  \label{sweyl}
\end{equation}
The transformations $\sigma _{i}$ are called simple Weyl reflections and
generate the Weyl group $W$. Their action on simple roots is described in
terms of the Cartan matrix associated with $\frak{g}$, 
\begin{equation*}
\alpha _{j}\rightarrow \sigma _{j}\alpha _{i}:=\alpha _{i}-A_{ij}\alpha
_{j}\;,\quad A_{ij}=\frac{2\left\langle \alpha _{i},\alpha _{j}\right\rangle 
}{\left\langle \alpha _{j},\alpha _{j}\right\rangle }\;.
\end{equation*}
There exists a longest element in $W$ in the sense that it is built up from
a maximal number of simple Weyl reflections. It is called the Coxeter
element or transformation\textbf{\ }and is defined by the product over all
simple Weyl reflections, $\sigma =\sigma _{1}\sigma _{2}\cdots \sigma _{n}$.
Clearly this definition depends on the ordering of the reflections, which is
a matter of choice. The Coxeter element is therefore only defined up to
conjugacy within the Weyl group. Following \cite{FLO} a unique Coxeter
element can be determined by introducing the concept of bicolouration for
Dynkin diagrams. To every vertex in the Dynkin diagram we assign a colour $%
c_{j}=\pm 1$ such that two vertices linked to each other are differently
coloured. This bicolouration polarizes the index set $\Delta =\{1,...,n\}$
into two subsets $\Delta _{\pm }$ and allows unambiguous specification of
the following Coxeter element: 
\begin{equation}
\sigma :=\sigma _{-}\sigma _{+},\quad \quad \sigma _{\pm }:=\prod_{i\in
\Delta _{\pm }}\sigma _{i}\;.  \label{Cox}
\end{equation}
Any Coxeter element shares the following properties \cite{kos}:

\begin{itemize}
\item[(C1)]  \emph{The Coxeter element fixes no non-zero vector.}

\item[(C2)]  \emph{It is of finite order, }$\sigma ^{h}=1,$\emph{\ where }$h$%
\emph{\ is the Coxeter number defined as the sum over the Kac or Dynkin
labels, } 
\begin{equation}
h:=1+\sum_{i=1}^{\text{rank\thinspace }\frak{g}}n_{i}\;.  \label{Coxnumber}
\end{equation}
\emph{Thus, the Coxeter element permutes the roots in orbits of length }$h$%
\emph{.}

\item[(C3)]  \emph{The eigenvalues of }$\sigma $\emph{\ are of the form} 
\begin{equation*}
\exp \frac{i\pi s_{j}}{h}\;,\quad j=1,...,\text{rank\thinspace }\frak{g}
\end{equation*}
\emph{where the characteristic set of integers }$1=s_{1}\leq s_{2}\leq
...\leq s_{n}=h-1$\emph{\ are called the exponents of the Lie algebra }$%
\frak{g}$\emph{\ satisfying the relation }$s_{n+1-i}=h-s_{i}$\emph{.}
\end{itemize}

The action of the Coxeter element on the root system splits the latter into
orbits. A convenient choice of representatives of these orbits are the
,,coloured'' simple roots, $\gamma _{i}=c_{i}\alpha _{i}$. The Coxeter
orbits $\Omega _{i}$ defined as 
\begin{equation}
\Omega _{i}:=\{\sigma ^{x}\gamma _{i}:1\leq x\leq h\}  \label{orbit}
\end{equation}
satisfy the crucial property that they do not intersect, i.e. $\Omega
_{i}\cap \Omega _{j}=\emptyset $, and are exhaustive on the set of roots.
Moreover, all $\gamma _{i}$'s lie in different orbits and all elements in
one orbit are linearly independent \cite{FLO}. Thus, the coloured simple
roots constitute a complete set of representatives for the Coxeter orbits $%
\Omega _{i}$.

\end{document}